\documentclass[journal]{IEEEtran}
\usepackage{amsmath,amsfonts}
\usepackage{graphicx}
\usepackage{amssymb}
\usepackage{siunitx}
\usepackage{cite}
\usepackage{mathtools}
\usepackage{bbold}
\usepackage{steinmetz}
\usepackage{stmaryrd}
\usepackage{bbm, dsfont}
\usepackage{amsthm}
\usepackage{verbatim}
\usepackage[version=4]{mhchem}
\usepackage[latin1]{inputenc}
\usepackage{tikz}
\usepackage{textcomp}
\usepackage{enumerate}
\usepackage{color}
\usepackage{pgfplots}
\usetikzlibrary{calc}
\usepackage{mathrsfs}
\usepackage{upgreek}
\usepackage{steinmetz}
\usepackage[outdir=./]{epstopdf}
\usepackage{lipsum}
\usepackage{amsbsy}
\usepackage{bm}
\usetikzlibrary{positioning}% To get more advances positioning options
\usetikzlibrary{shapes,arrows}
\usepackage{multirow}

\pgfplotsset{
	every tick label/.append style={scale=1},
	every axis/.append style={
	}
}

\pgfplotsset{
	grid style = {
		dash pattern = on 0.05mm off 1mm,
		line cap = round,
		black,
		line width = 0.5pt
	}
}
\usetikzlibrary{shapes.multipart,intersections}

\newcommand{\tx}{\textnormal}

\newsavebox\myboxA
\newsavebox\myboxB
\newlength\mylenA

\newcommand*\xoverline[2][0.75]{%
	\sbox{\myboxA}{$\m@th#2$}%
	\setbox\myboxB\null% Phantom box
	\ht\myboxB=\ht\myboxA%
	\dp\myboxB=\dp\myboxA%
	\wd\myboxB=#1\wd\myboxA% Scale phantom
	\sbox\myboxB{$\m@th\overline{\copy\myboxB}$}%  Overlined phantom
	\setlength\mylenA{\the\wd\myboxA}%   calc width diff
	\addtolength\mylenA{-\the\wd\myboxB}%
	\ifdim\wd\myboxB<\wd\myboxA%
	\rlap{\hskip 0.5\mylenA\usebox\myboxB}{\usebox\myboxA}%
	\else
	\hskip -0.5\mylenA\rlap{\usebox\myboxA}{\hskip 0.5\mylenA\usebox\myboxB}%
	\fi}
\makeatother

\usepackage{subfigure} % For subfigures
\usepackage{graphicx}

\begin{document}
\bstctlcite{IEEEexample:BSTcontrol}

\title{Optimal Thermal Management and Charging of Battery Electric Vehicles over Long Trips}

%\title{Eco-driving and Optimal Battery Thermal Management of Battery Electric Vehicles Driving in a Cold Climate}

\author{Ahad Hamednia, Victor Hanson, Jiaming Zhao, Nikolce Murgovski, Jimmy Forsman, Mitra Pourabdollah, Viktor Larsson, and Jonas Fredriksson  % <-this % stops a space
\thanks{N. Murgovski and Jonas Fredriksson are with the Department of Electrical Engineering, Chalmers University of Technology,
Gothenburg 412 96, Sweden.}% <-this % stops a space
\thanks{A. Hamednia, V. Hanson, J. Forsman, M Pourabdollah, and Viktor Larsson are with the Department of Vehicle Energy and Motion Control, and J. Zhao is with the Exterior Systems team, Volvo Car Corporation, Gothenburg 405 31, Sweden (e-mail: ahad.hamednia@volvocars.com).}}
\maketitle

\begin{abstract}
This paper studies optimal thermal management and charging of a battery electric vehicle driving over long distance trips. The focus is on the potential benefits of including a heat pump in the thermal management system for waste heat recovery, and charging point planning, in a way to achieve optimality in time, energy, or their trade-off. An optimal control problem is formulated, in which the objective function includes the energy delivered by the charger(s), and the total charging time including the actual charging time and the detour time to and from the charging stop. To reduce the computational complexity, the formulated problem is then transformed into a hybrid dynamical system, where charging dynamics are modelled in the domain of normalized charging time. Driving dynamics can be modelled in either of the trip time or travel distance domains, as the vehicle speed is assumed to be known a priori, and the vehicle is only stopping at charging locations. Within the hybrid dynamical system, a binary variable is introduced for each charging location, in order to decide to use or skip a charger. This problem is solved numerically, and simulations are performed to evaluate the performance in terms of energy efficiency and time. The simulation results indicate that the time required for charging and total energy consumption are reduced up to \SI{30.6}{\%} and \SI{19.4}{\%}, respectively, by applying the proposed algorithm.
\end{abstract}

\begin{IEEEkeywords}
Grid-to-meter energy efficiency, thermal management, charging, heat pump, charge point planning
\end{IEEEkeywords}

\IEEEpeerreviewmaketitle

\section{Introduction}\label{sec:intro}
\IEEEPARstart{R}{ecently} electric vehicles (EVs) have gained considerable attention among researchers, manufacturers, and users, due to their advanced and sustainable technologies for counteracting drawbacks of conventional vehicles, e.g. limited fuel resources, severe environmental impact, and high maintenance and operating costs~\cite{kumar20}. Accordingly, the EV market has grown rapidly over the last few years, and several car companies have stated that they will only produce electric vehicles in the near future~\cite{andwari17}. In particular, battery electric vehicles (BEVs) are identified as a promising choice for achieving the decarbonized light-duty vehicle fleet. However, there still exist several challenges impeding the widespread deployment of BEVs, mostly related to energy cost, limited driving range, charging time, and thermal management. These issues become even more important to consider when planning for long-distance trips, i.e. exceeding the vehicle's range~\cite{sanguesa21}.

Although the range can vary over a large distance window~\cite{suarez19}, still the majority of cost-effective BEV models fail to fully meet the range requirement of long trips, highlighting the significance of reducing total energy consumption as well as improving fast charging technology, for higher customer acceptance of BEVs. Lately, a high-power fast charging technology has been introduced, aiming at recharging a battery up to \SI{80}{\%} state of charge (SoC) within \SI{15}{min}, in order to provide more convenient long-trip experiences~\cite{mahfouz19}.

Apart from the charger's rated power, the charging time is also highly influenced by the fast charging properties of the battery. This is mainly characterised by the battery's chemistry, SoC, temperature, and health state, which may negatively affect the charging rate~\cite{jaguemont18}. Thus, solutions associated with the BEV's fast charging are required to incorporate various aspects rather than just focusing on increasing the maximum power provided by the charger~\cite{keyser17,rafi20}.  

One crucial factor that can significantly improve charging time, total energy consumption, and passenger comfort, especially in harsh climates, is to develop an adequate thermal management (TM)~\cite{jaguemont15,zhang17}. Lithium-ion (Li-ion) batteries, known as a widely used alternative in the market, are highly temperature sensitive~\cite{wang18}. Excessive battery temperatures can cause corrosion and even explosion by creating bubbles, bulges, sparks, and flames~\cite{hannan18}. Furthermore, at sub-zero Celsius temperatures, the battery performance is severely deteriorated due to a considerably slowed electrochemical process within the battery cells~\cite{wu20,wang22}. This yields a severe reduction in the cell's available power and energy, thereby significantly increasing the charging time~\cite{jeffs19}. Moreover, to minimize the total energy consumption of the vehicle, it is essential to incorporate the TM when optimising the grid-to-meter energy efficiency of the BEV~\cite{amini19,chen20,hu20}. 
In this context, several research works have been conducted, mainly by formulating an optimal control problem (OCP) that can be solved by different optimization tools.

\begin{figure*}[t!]
 \centering
 \includegraphics[width=.95\linewidth]{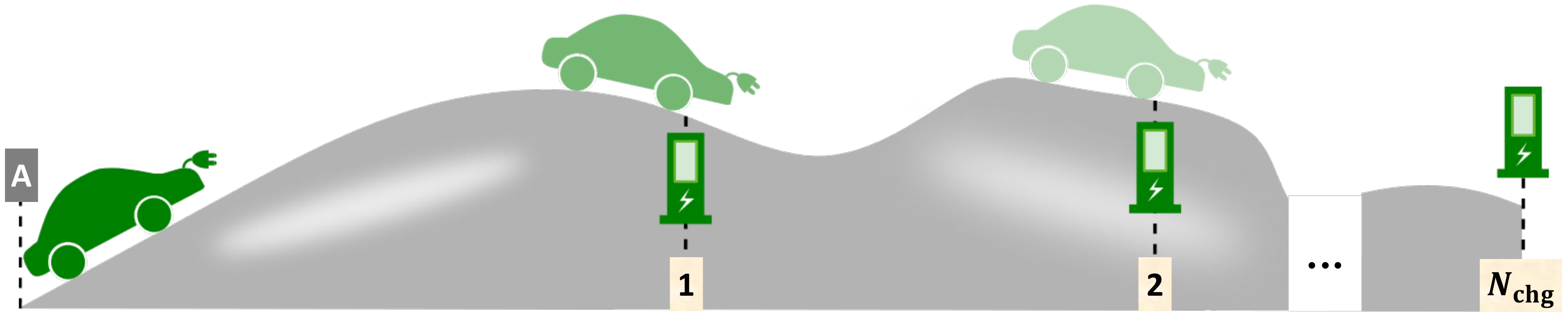}
  \caption{\footnotesize A BEV starts its trip from point A, and drives in hilly terrain. The indices $1$, $2$,\dots represent the charging stations, and $N_\tx{chg}$ denotes the total number of charging locations.}
  \label{fig:scenario}
\end{figure*}

Dynamic programming (DP)~\cite{bellman57} is used in~\cite{jaguemont16} for developing an algorithm for the TM of a vehicle that is unplugged from the electrical grid and parked outside at a low ambient temperature. The goal of this study is to find an optimal trade-off between contained energy in the battery pack, and the cell degradation of being exposed to cold weather. However, the main disadvantage of the DP approach is expressed as the \textit{curse of dimensionality}, which refers to the exponential growth of computational time with the dimension of the OCP. As an alternative approach, Pontryagin's Maximum Principle (PMP)~\cite{pontryagin62} is applied in~\cite{hartl95}, for maximising the expected battery life with minimum energy consumed. PMP suggests a way to reduce the computational complexity of the high-dimensional optimisation problems, by adjoining system dynamics to the objective function and neglecting constraints on state variables. Furthermore, several TM strategies have been proposed using Model Predictive Control (MPC) scheme for increasing energy efficiency via optimal cooling/heating~\cite{zhu18,lopez16,lopez17}. Moreover, the TM is studied for vehicles with a given drive cycle~\cite{chen2020joint}, or with future speed prediction, to be incorporated into the energy efficiency analysis~\cite{amini19}. In the context of the TM of electrified vehicles, several research efforts have been carried out with a focus on waste heat recovery (WHR)~\cite{rodriguez19}, referred to as an  energy recovery process by transferring heat from one part to another part within the vehicle and, thus, improve the energy efficiency. In~\cite{lee22}, a multi-level WHR system with an improved heat transfer capacity is developed, where the battery temperature is maintained within an appropriate range. Also, a novel HP system is designed for electric buses in~\cite{han20}, where the heating performance of the TM system is enhanced in cold environments. Despite the contributions provided by developing numerous TM strategies for vehicles, the technical literature lacks investigation on joint optimal charging and TM over long trips, with a WHR ability and charge point planning.

As an extension to our earlier work~\cite{hamednia2022a}, this paper addresses a BEV driving on a road with hilly terrain. The vehicle's travelled distance is greater than its range; there is thus a need for at least one charging stop along the route. In this paper, the following goals are addressed:
\begin{itemize}
    \item Develop an algorithm to achieve optimal charging and TM of a BEV on long trips, capturing both driving and charging modes of the vehicle.
    \item Quantify the trade-off between charging time and energy efficiency.
    \item Investigate the benefits of including a heat pump (HP) in the TM system for WHR.
    \item Plan the charging locations, in favour of obtaining optimality in time, energy, or their trade-off.
\end{itemize}

To achieve the above-mentioned goals, an OCP is formulated for charging and TM of a BEV. The objective of the OCP is to find the optimal compromise between the energy delivered by the charger(s), and the \textit{total charging time} referred to as the actual charging time and the detour time to and from the charging locations. The TM system includes an HP, a high-voltage coolant heater (HVCH), and heating, ventilation, and air conditioning (HVAC). HP is used for the WHR purposes, and HVCH and HVAC are employed, respectively for heating and cooling of the battery and cabin. The driving dynamics can be described in either of the space or trip time domains. However, charging dynamics is modelled in terms of normalized charging time. Thus, the OCP transforms into a hybrid dynamical system (HDS). Note that the actual charging time is treated as a scalar variable, which is optimized simultaneously with the optimal state and control trajectories that belong to the driving and charging modes. Also, for each charging location, a binary variable is defined to optimally plan the charging stops, in favour of further optimising the energy efficiency and/or trip time. Such formulation procedure turns the HDS into a mixed-integer optimisation problem.

The rest of the paper is outlined as follows. In Section~\ref{sec:model}, electrical and thermal modelling of the electric powertrain are addressed. Section~\ref{sec:pbpgridlim} illustrates the constraints on the battery and grid power values. In Section~\ref{sec:method}, the HDS is formulated, covering the vehicle's operation during both driving and charging modes. In Section~\ref{sec:res} simulation results are presented. Section~\ref{sec:dis} discusses the obtained results. Finally, Section~\ref{sec:con} includes the conclusion of the paper and suggestions regarding possible future research directions.

% \vspace{5cm}
% Among all parts of the electric powertrain, electric battery is the key component because:
% \begin{itemize}
%     \item The battery is the most expensive electrical component of the powertrain.
%     \item The vehicle's electric range relies almost entirely on the battery.
%     \item The battery is a heavy component, i.e. it can weigh up to about $\SI{500}{kg}$, hence it has a large thermal mass.
%     \item It is expensive to actively heat/cool the battery.
%     \item The time constant of battery temperature rate while the heating/cooling can be very long.
% \end{itemize}
% Thus, it is crucial to model the battery to study its electric and thermal behaviors.
% \vspace{1cm}
% Electric battery is the key component among all parts of the electric powertrain, because: (i) the battery is the most expensive electrical component of the powertrain; (ii) the vehicle's electric range relies almost entirely on the battery; (iii) the battery is a heavy component, i.e. it can weigh up to about $\SI{500}{kg}$, hence it has a large thermal mass; (iv) the time constant of battery temperature rate while the heating/cooling can be very long.

\section{Modelling}\label{sec:model}
This section addresses the vehicle driving mission and a multi-domain configuration of an electric powertrain, describing the connection of the powertrain components via electrical, thermal, and mechanical paths.

\subsection{Vehicle driving mission}\label{subsec:mission}
Consider a BEV that starts its trip from point A, and drives in hilly terrain, as depicted in Fig.~\ref{fig:scenario}. As the vehicle moves forward, the battery is depleted. The battery temperature may be adjusted by different heating/cooling sources within the powertrain. Along the driving route, multiple charging possibilities are considered, as the vehicle's trip length is greater than its range. In realistic driving situations, it is preferable to plan the charging stops, to achieve optimal trip time and/or charging cost.

In this paper, we assume the vehicle speed to be known a priori, in which the vehicle stops only during charging (and not during driving). This allows us to identically formulate the driving dynamics, in either space or trip time domains, without adding any complexity to the algorithm developed later in Section~\ref{sec:method}. Here, we freely choose the spatial domain to associate the system trajectories with space-defined events, such as speed limits and charging locations. Thus, the vehicle's driving time, $t$, is calculated by integrating the vehicle speed, as
\begin{align}
    t(s) = \int_{0}^s\frac{\tx{d}x}{v(x)},\label{eq:time_dyn_t}
\end{align} 
where $s$ and $v$ denote travelled distance and the vehicle speed, respectively. The formulation of charging dynamics is postponed to Section~\ref{sec:method}.

\subsection{Multi-domain Powertrain Configuration}\label{subsec:powertrain}
\begin{figure}[t!]
 \centering
 \includegraphics[width=.9\linewidth]{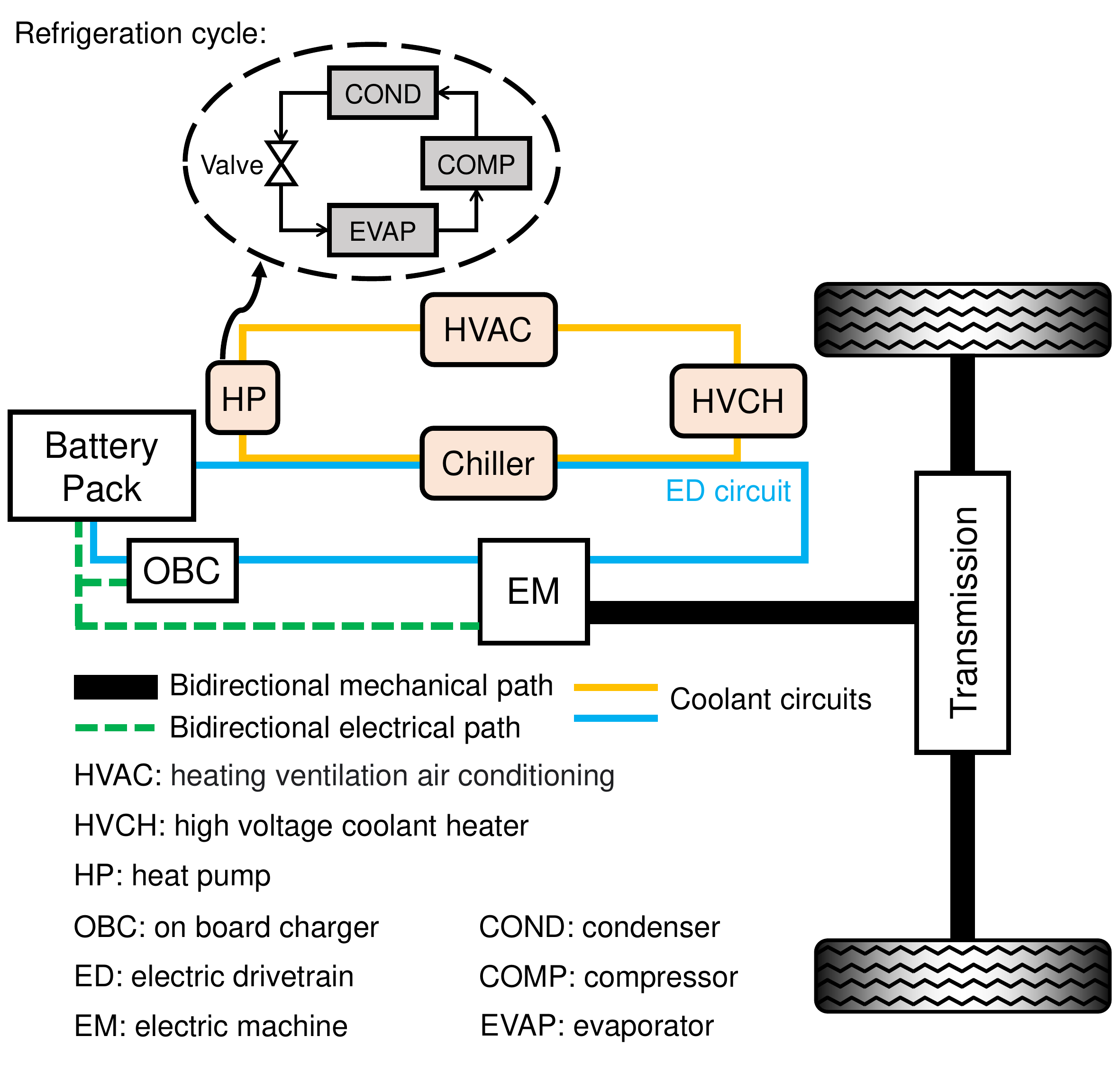}%\vspace{-.3cm}
  \caption{\footnotesize Schematic diagram of the studied electric powertrain, which consists of propulsion components, i.e. a  battery, an EM, and a  transmission system, an onboard charger, and a thermal management system. The thermal management system consists of HVCH, HVAC, and a heat pump, which are used for actively adjusting the battery pack and cabin compartment temperatures.}
  \label{fig:powertrain}
\end{figure}

A schematic diagram of the studied electric powertrain is demonstrated in Fig.~\ref{fig:powertrain}. The powertrain includes propulsion components, i.e. a battery for energy supply/storage, an electric machine (EM), and a transmission system. %Depending on the EM's operating mode, i.e. generating or motoring, the electric power flow through the electrical path is bidirectional. Accordingly, the electrical energy is stored in the battery during generating mode, or supplied to the EM during motoring mode. 
In addition to the propulsion components, the powertrain is equipped with an onboard charger (OBC), as a device to regulate the electricity flow from the electrical grid to the battery, monitor the charge rate, and protect the battery from over-current charging. Furthermore, the electric powertrain includes a thermal management system, comprising HVAC, HVCH, and HP. The HVAC and HVCH are mainly used, respectively for cooling and heating of the battery pack and cabin compartment. Also, an HP is generally employed for transferring heat from the heat source at low temperature, i.e. the battery, to heat sink at higher temperature, for e.g. the cabin compartment and/or ambient air. To achieve this, work is required, as heat cannot spontaneously flow from a colder place to a warmer location, according to the second law of thermodynamics~\cite{moran10}. As depicted in Fig.~\ref{fig:powertrain}, the operating principle of HPs can be summarized into a \textit{refrigeration cycle}, which consists of five major components: an evaporator, compressor, condenser, expansion valve, and refrigerant. Thus, the evaporator absorbs heat from the battery pack and turns the refrigerant from liquid mode into a low-pressure gas that is delivered to the compressor. Then the compressor pressurises the gas and dispatches it to the condenser. Later, the condenser cools down the hot gas, turns it into a liquid, and expels the extracted heat from the refrigerant to the cabin compartment and/or ambient air. Finally, the high-pressure liquid refrigerant departed from the condenser becomes a low-pressure liquid by passing through the expansion valve; and the cycle starts over again. The merit of an HP is specified by a parameter called the coefficient of performance (CoP), defined as a ratio of useful heating provided (for the cabin compartment) to the net work required, as
\begin{align}
    \tx{cop}(T_\tx{b}(s),P_\tx{hp}(s))=\frac{Q^\tx{b}_\tx{hp}(T_\tx{b}(s))+P_\tx{hp}(s)}{P_\tx{hp}(s)}\label{eq:cop},
\end{align}
where $T_\tx{b}$ is the battery pack's temperature, $P_\tx{hp}$ is the rate of the net work put into the cycle, and $Q^\tx{b}_\tx{hp}$ is rate of the heat removed from the battery pack and electric drivetrain (ED). Hereafter, $P_\tx{hp}$ is called HP power. The three domains of the powertrain configuration are elaborated in Sections~\ref{subsubsec:el_dom}-\ref{subsubsec:mech_dom}.

\subsubsection{Electrical Domain}\label{subsubsec:el_dom}
Depending on the EM's operating mode, i.e. generating or motoring, the electric power flow through the electrical path is bidirectional, as shown in Fig.~\ref{fig:powertrain}. Accordingly, electrical energy is charged to the battery during the generating mode, or supplied to the EM during the motoring mode. The battery is modelled using an equivalent circuit, which includes a voltage source $U_\tx{oc}$, known as open-circuit voltage, and an internal resistance $R_\tx{b}$. The open-circuit voltage is usually proportional to the battery SoC. Also, as the battery temperature is raised, the ions inside the battery cells get more energized, which results in reduced resistance against the ions' displacement. Thus, the internal resistance is commonly a nonlinear monotonically decreasing function of the battery temperature~\cite{zhu18}. Note that the slight mismatch between the internal resistance while charging and discharging is overlooked in this paper. The battery SoC dynamics is calculated by
\begin{align}
    \tx{soc}'(s)=-\frac{P_\tx{b}(s)}{C_\tx{b}U_\tx{oc}(\tx{soc}(s))v(s)},
    \label{eq:soc_dyn_s}
\end{align}
where $P_\tx{b}$ is battery power, including internal resistive losses, and $C_\tx{b}$ is maximum capacity of the battery pack. $P_\tx{b}$ is negative while charging, and is positive when discharging. Note that throughout this paper, $x'$ represents the space derivative of an arbitrary variable $x$, i.e. $x'=\tx{d}x/\tx{d}s$.

\begin{figure*}[t!]
\centering
\subfigure[Battery discharge power limit.]{
 \includegraphics[width=.425\linewidth]{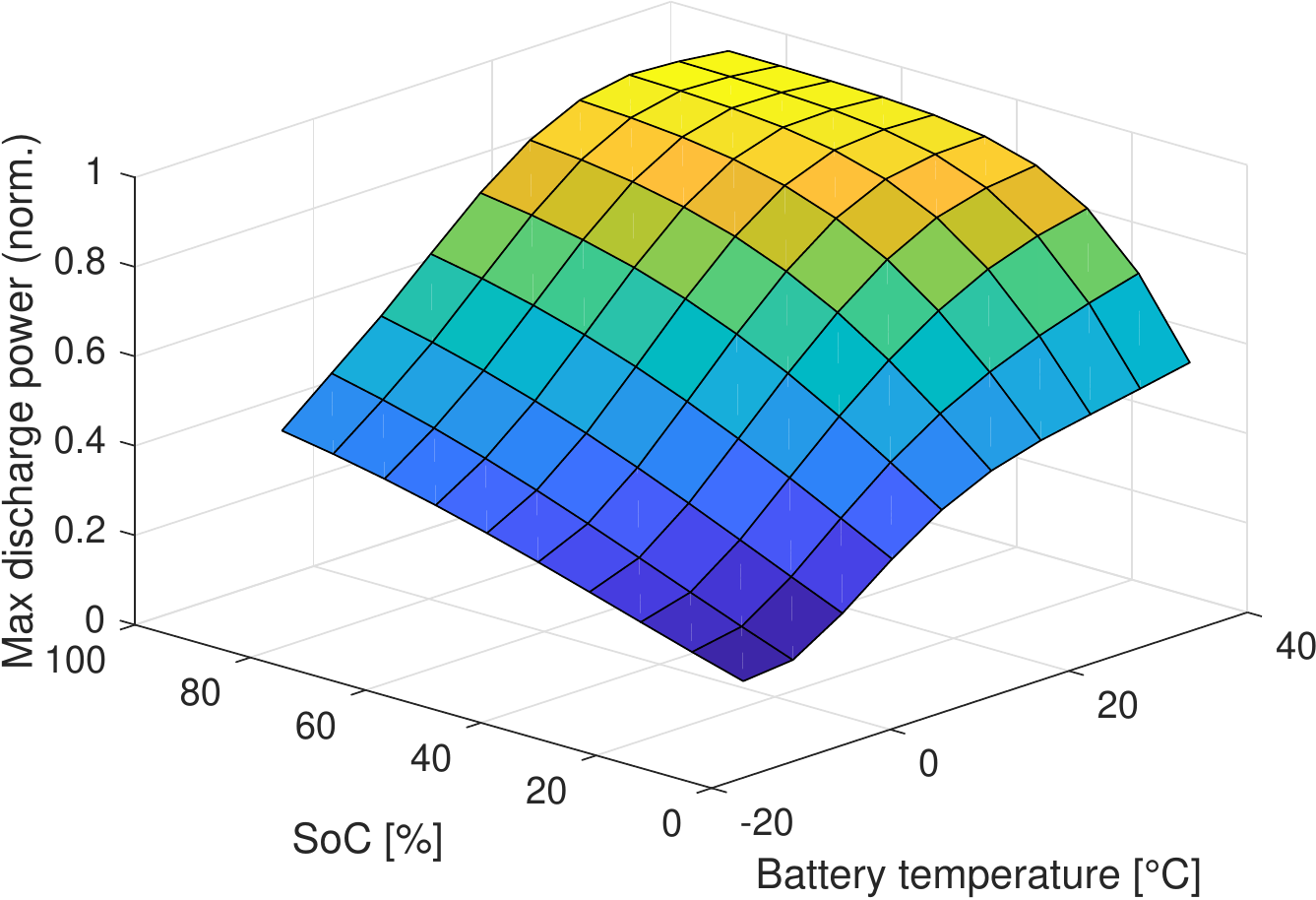}\hspace{1cm}
\label{fig:pbdchglim}
}
\subfigure[Battery charge power limit.]{

 \includegraphics[width=.425\linewidth]{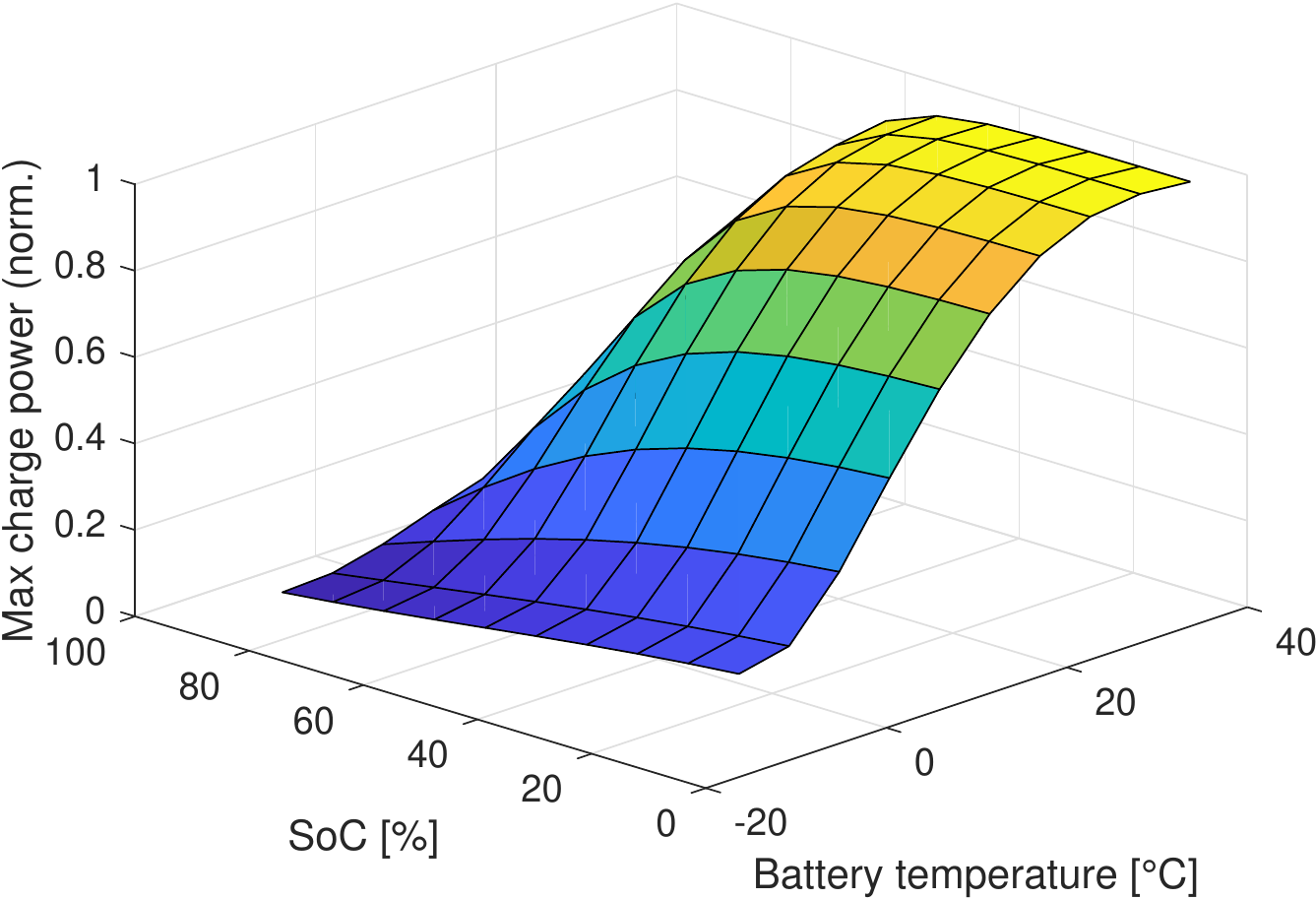}
\label{fig:pbchglim}
}
\caption{Normalised absolute value of discharge and charge power bounds versus battery temperature and SoC for the studied battery in this paper.}
\label{fig:pblim}
\end{figure*}

\subsubsection{Thermal Domain}\label{subsubsec:th_dom}
According to the fundamental thermodynamic principle~\cite{moran10}, the changing rate of the battery pack's temperature $T_\tx{b}$ is modelled using a lumped-parameter thermal model, as 
\begin{align}
T_\tx{b}'(s)=\frac{1}{c_\tx{p}m_\tx{b}v(s)}\big(Q_\tx{pass}(\cdot)+Q_\tx{act}(\cdot)+Q_\tx{exh}(\cdot)\big),\label{eq:tb_dyn_s}
\end{align}
where $c_\tx{p}$ and $m_\tx{b}$ are the battery pack's specific heat capacity and total mass, respectively, $Q_\tx{pass}$ is the rate of induced heat by passive heat sources affecting the battery temperature, $Q_\tx{act}$ is the active heat rate from or removed by components, e.g. HVAC, HVCH, and HP, that can actively affect the battery pack temperature, $Q_\tx{exh}$ is rate of the heat exchanged among the battery pack, ambient air and/or the chassis of the vehicle, and the symbol $\cdot$ is a compact notation used for expressing multiple variables of a function. Note that the nonuniform distribution of the battery pack temperature due to heat diffusion is neglected in this paper, which reduces the complexity of the thermal model. Accordingly, crust and core battery pack temperatures are assumed to be identical. 

The passive heat generation rate
\begin{align}
\begin{split}
&Q_\tx{pass}(\tx{soc}(s),T_\tx{b}(s),v(s),a_\tx{t}(s))=\\
&\hspace{2cm}R_\tx{b}(T_\tx{b}(s))\frac{P_\tx{b}^2(s)}{U_\tx{oc}^2(\tx{soc}(s))}+Q_\tx{ed}(v(s),a_\tx{t}(s))\label{eq:qrqed},
\end{split}    
\end{align}
includes 1) the produced heat due to the battery's internal resistive losses, referred to as \textit{irreversible ohmic Joule heat}; and 2) the heat produced by the ED power losses, $Q_\tx{ed}$, which is dependent on the vehicle speed and traction acceleration $a_\tx{t}$.

The active heat generation rate 
\begin{align}
\begin{split}
&Q_\tx{act}(P_\tx{hvch}^\tx{b}(s),P_\tx{hvac}^\tx{b}(s),P_\tx{hp}(s))=\eta_\tx{hvch}P_\tx{hvch}^\tx{b}(s)\\
&\hspace{2.5cm}-\eta_\tx{hvac}P_\tx{hvac}^\tx{b}(s)-Q_\tx{hp}(s) \label{eq:qact}.
\end{split}    
\end{align}
includes HVCH power conversion for heating the battery pack, $P_\tx{hvch}^\tx{b}$, with efficiency of $\eta_\tx{hvch}$, HVAC power conversion for cooling the battery pack, $P_\tx{hvac}^\tx{b}$, with efficiency of $\eta_\tx{hvac}$, and rate of the heat removed from the battery pack by HP.

The convective heat exchange rate between the battery pack and ambient air depends on the ambient temperature $T_\tx{amb}$, battery temperature, and vehicle speed, as
\begin{align}
    Q_\tx{exh}(T_\tx{amb}(s),T_\tx{b}(s),v(s))=\gamma(v(s))(T_\tx{amb}(s)-T_\tx{b}(s)),\label{eq:qexh}
\end{align}
where $\gamma>0$ is a speed-dependent coefficient of heat exchange.

\subsubsection{Mechanical Domain}\label{subsubsec:mech_dom}
Similar to the electrical path, the mechanical path is also bidirectional, as depicted in Fig.~\ref{fig:powertrain}. The EM when operated in motoring mode, provides propulsion power, which is delivered to the wheels through the mechanical path via the transmission system. Thus, the EM rotational speed and output torque are translated by the transmission system into vehicle speed and traction acceleration, respectively. Furthermore, the EM when operated in generating mode, transforms the vehicle's kinetic energy at the wheels via the mechanical path into electrical energy to be stored in the battery.

\section{Bounds on Battery and Grid Power Values}\label{sec:pbpgridlim}
The bounds on available battery power during discharging and charging are formulated as functions of battery temperature and SoC as
\begin{align}\label{eq:pbbound}
\resizebox{0.96\hsize}{!}{$P_\tx{b}(s)\in \begin{cases}
	[P_\tx{b,chg}^{\min}(\tx{soc}(s),T_\tx{b}(s)),P_\tx{b,dchg}^{\max}(\tx{soc}(s),T_\tx{b}(s))], & \text{$s\in \mathcal{S}_\tx{drv}$}\vspace{0.25cm}\\
    [\zeta_iP_\tx{b,chg}^{\min}(\tx{soc}(s),T_\tx{b}(s)),0], & \text{$s\in \mathcal{S}_\tx{chg}^{i}$}
\end{cases}$}
\end{align}
where $P_\tx{b,dchg}^{\max}>0$ and $P_\tx{b,chg}^{\min}<0$ are the bounds on the battery discharge and charge power, respectively, $i\in\mathcal{I}=\{1,2,\dots,N_\tx{chg}\}$ is charger index, $N_\tx{chg}$ is total number of charging locations along the driving route, and $\mathcal{S}_\tx{drv}$ and $\mathcal{S}_\tx{chg}$ represent sets of driving and charging distance instances, respectively. Also, $\zeta\in\mathbb{Z}=\{0,1\}$ is a binary variable defined for each charging location, in order to decide whether to skip the charger, $\zeta=0$, or use it, $\zeta=1$. Note that $P_\tx{b,chg}^{\min}$ may differ in driving and charging modes, whereas it is here assumed that the same bound is imposed for simplicity, and without loss of generality. The negative power limit during driving is due to regenerative braking, within which the kinetic energy at the wheels is transformed into electrical energy to be stored in the battery. As demonstrated in Fig.~\ref{fig:pbdchglim}, the studied battery discharge power limit is proportional to both the battery temperature and SoC level. However, the battery charge power limit is proportional to the battery temperature and inverse of SoC level, as depicted in Fig.~\ref{fig:pbchglim}.

For $i\in\mathcal{I}$ and $\zeta\in\mathbb{Z}$, the bound on the $i$th charger's provided power is given by
\begin{align}
P^{i}_\tx{grid}(s) \in \begin{cases}
	\{0\}, & \text{$s\in \mathcal{S}_\tx{drv}$},\vspace{0.25cm}\\
   [0,\zeta_iP^{i,\max}_{\tx{grid}}], & \text{$s\in \mathcal{S}_\tx{chg}^{i}$}
\end{cases}\label{eq:pgridbound}
\end{align}
where $P_\tx{grid}^{i,\max}$ is rated power of the $i$th charger. Although it is assumed that the vehicle power demand is not supplied by the grid power during the driving mode, it is possible to do so on a road with charging lanes installed~\cite{limb18}, by directly applying a method developed earlier in~\cite{hamednia2022a} in combination with the method provided later in Section~\ref{sec:method}.

Considering the battery and grid power limits \eqref{eq:pbbound} and \eqref{eq:pgridbound}, the power balance equation can be written as
\begin{align}
\begin{split}
&P^{i}_\tx{grid}(s)+P_\tx{b}(s)=R(T_\tx{b}(s))\frac{P_\tx{b}^2(s)}{U_{\tx{oc}}^2(\tx{soc}(s))}+P_\tx{prop}(v(s),a_\tx{t}(s))\\
&+P^\tx{b}_\tx{hvch}(s)+P^\tx{b}_\tx{hvac}(s)+P^{\tx{c}}_{\tx{hvch}}(T_\tx{amb}(s))+P_\tx{hp}(s)+P_\tx{aux}(s),
\end{split}\label{eq:p_balance}    
\end{align}
where $P_\tx{prop}$ is propulsion power including the internal powertrain losses, $P^{\tx{c}}_{\tx{hvch}}$ is the HVCH power consumed for heating the cabin compartment, and $P_\tx{aux}$ is auxiliary power demand used for lights, infotainment, etc.

System state variables and control inputs can be stacked into state and control vectors, respectively $\tx{x}$ and $\tx{u}$, as
\begin{align}
    \tx{x}(s)=\begin{bmatrix} \tx{soc}(s)\\T_\tx{b}(s)\end{bmatrix}, \ \tx{u}(s)=\begin{bmatrix} P^\tx{b}_\tx{hvch}(s)\vspace{0.15cm}\\ P^\tx{b}_\tx{hvac}(s)\vspace{0.15cm}\\P^\tx{hp}_\tx{b}(s)\vspace{0.15cm}\\
    P_\tx{grid}(s) \end{bmatrix}.
\end{align}
Thus, according to \eqref{eq:soc_dyn_s} and \eqref{eq:tb_dyn_s}, the governing dynamics describing the battery SoC and temperature variations in the spatial domain can be summarized as
\begin{align}
    \frac{\tx{d}\tx{x}(s)}{\tx{d}s}=\frac{1}{v(s)}h(\tx{x}(s),\tx{u}(s),s),\label{eq:h}
\end{align}
with $h$ defined as a vector function.

\section{Problem Statement}\label{sec:method}
\begin{figure}[t!]
 \centering
 \includegraphics[width=\linewidth]{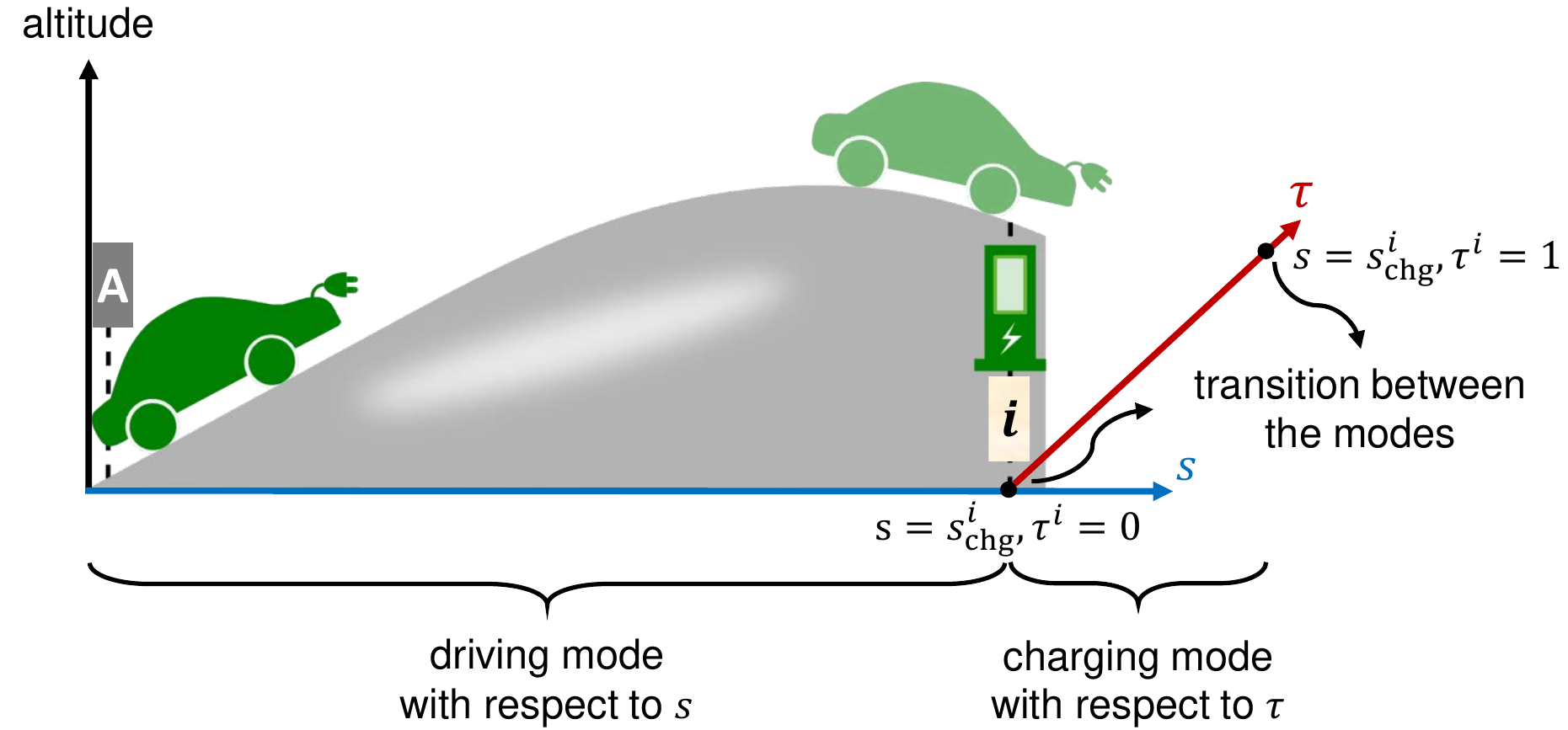}
  \caption{\footnotesize Hybrid dynamical system demonstration including driving mode, charging mode and transition between these two modes. During the driving and charging modes decisions are planned with respect to $s$ and $\tau^i$, $i\in\mathcal{I}$, respectively.}
  \label{fig:module}
\end{figure}

Despite the vehicle's fixed position at the charging stop, there will still be dynamic variations in the battery temperature and SoC while charging. Thus, to find the optimal trade-off between time and energy cost during both the driving and charging modes, it is not possible to formulate a single optimisation problem, within which decisions are always made with respect to $s$. Subsequently, we propose modelling of the charging dynamics in a temporal domain, where
decisions are planned along a normalized charging time, $\tau^i \in [0,1]$, defined, as
\begin{align}
\tau^i=\frac{t}{t_\tx{chg}^{i}}, \quad t\in \mathcal{T}_\tx{chg}^{i}, i\in\mathcal{I},\label{eq:t2tau}
\end{align}
where $t$ is trip time, and $\mathcal{T}_\tx{chg}^{i}$ and $t_\tx{chg}^i$ denote respectively a set of charging time instants and charging time, at the $i^{th}$ charging station. Thus, by choosing a distinct independent variable describing each mode, i.e. $s$ for the driving mode and $\tau^i$ for the charging modes, as well as considering the binary variable $\zeta$, a mixed-integer HDS can be formulated. A demonstration of the HDS including the driving and charging modes as well as the transition between the modes is shown in Fig.~\ref{fig:module}, whereby repeating such a combination, it is possible to incorporate multiple charging locations within the problem. 

Following \eqref{eq:t2tau} and the derivative chain rule, the relation between the space derivative and the derivative with respect to $\tau^{i}\in[0,1]$, $i\in\mathcal{I}$ is given by
\begin{align}
    \frac{\tx{d}\tx{x}}{\tx{d}s}=\frac{\tx{d}\tx{x}}{\tx{d}\tau}\frac{1}{t_\tx{chg}v(s)},\label{eq:s2tau}
\end{align}
where $\frac{1}{t_\tx{chg}v(s)}=\frac{\tx{d}\tau}{\tx{d}t}\frac{\tx{d}t}{\tx{d}s}$. Hereafter, the variables notated with subscripts/superscripts `drv' or `chg', correspond to the previously introduced variables that now belong specifically to the driving mode or charging mode, respectively.
%Note that state variables, control inputs and governing dynamics describing each mode may differ with those from the other mode's.
Note that the charging cost can be defined as the cost of electrical energy provided by the charger and/or the time spent occupying the charging spot, depending on the pricing policy of a charger.

\subsection{Objective Function}\label{subsec:obj}
The objective function of the optimisation problem is defined as
\begin{align}
\begin{split}
&J(\cdot)=\sum_{i=1}^{N_\tx{chg}}\Big(\int_{0}^{1}
c^{i}_\tx{e}P^{i}_\tx{grid}(\tau^i)\tx{d}\tau^i+c_\tx{t,chg}t_\tx{chg}^{i}\\
&\hspace{2.5cm}+c^{i}_\tx{occ}\max\big(0,t^{i}_\tx{chg}-t_\tx{occ}^{i}\big)+c_\zeta\zeta_i\Big),
\end{split}\label{eq:J}
\end{align}
where $J$ includes
\begin{itemize}
   \item a charger's supplied electrical energy to the vehicle, %as  
%   \begin{align}
%       \int_{0}^{1}c^{i}_\tx{e}P^{i}_\tx{grid}(\tau^i)\tx{d}\tau^i,\label{eq:J_echg}
%   \end{align}
   where $c_\tx{e}$ denotes currency per-kilowatt-hour cost of the charged energy.
   \item a penalty on charging time with $c_\tx{t,chg}$ as the penalty coefficient.
   %as
%     \begin{align}
%     \sum_{i=1}^{N_\tx{chg}}c_\tx{t,chg}t_\tx{chg}^{i},\label{eq:J_chgtime}
%   \end{align}
%   where $c_\tx{t,chg}$ is the penalty coefficient. %\red{Note that the driving time also includes the detour time that is spent for driving from the main route to a charging station and back.}
   
   \item a cost of occupying the charger for longer time than $t_\tx{occ}\geq 0$, 
%   as
%     \begin{align}
%       \sum_{i=1}^{N_\tx{chg}}c^{i}_\tx{occ}\max\big(0,t^{i}_\tx{chg}-t_\tx{occ}^{i}\big),\label{eq:J_occuchg}
%   \end{align}
   where $c_\tx{occ}$ is currency per-minute cost, and a scalar variable $t_\tx{chg}$ represents the charging time. Note that with non-zero value of $c_\tx{T}$, the charging time is penalized twice, due to an occupied charger and/or a longer charging time.
   
   \item a detour cost to penalise the number of charging occasions, 
%   as
%     \begin{align}
%       \sum_{i=1}^{N_\tx{chg}}c_\zeta\zeta_i,\label{eq:J_chgnoopt}
%   \end{align}   
   where $c_\zeta$ is the penalty factor.
\end{itemize}

\subsection{Mixed-integer Hybrid Dynamical System Formulation}\label{subsec:pf_hybrid}
Using \eqref{eq:h} and \eqref{eq:s2tau}, the mixed-integer HDS formulation for $i\in\mathcal{I}$, $\tau^i\in[0,1]$, and $\zeta\in \mathbb{Z}$ can now be summarized as
{\allowdisplaybreaks
\begin{subequations} \label{eq:phyb}
\begin{align}
&\min_{\tx{u}_\tx{drv}(s),\tx{u}^{i}_\tx{chg}(\tau^i),t^i_\tx{chg},\zeta_i}J(\cdot)\\
&\text{subject to:} \nonumber\\
&\frac{\tx{d}\tx{x}_\tx{drv}(s)}{\tx{d}s}=\frac{1}{v(s)}h(\tx{x}_\tx{drv}(s),\tx{u}_\tx{drv}(s),s), \quad s\in\mathcal{S}_\tx{drv}\label{eq:phyb_sdyn}\\
&\frac{\tx{d}\tx{x}^{i}_\tx{chg}(\tau^i)}{\tx{d}\tau^i}=t^{i}_\tx{chg}h(\tx{x}^{i}_\tx{chg}(\tau^i),\tx{u}^{i}_\tx{chg}(\tau^i),\tau^i), \quad s\in s^{i}_\tx{chg}\label{eq:phyb_taudyn}\\
&g_\tx{drv}(\tx{x}_\tx{drv}(s),\tx{u}_\tx{drv}(s),s)\leq 0, \quad s\in\mathcal{S}_\tx{drv}\label{eq:phyb_gscns}\\
&g_\tx{chg}(\tx{x}^{i}_\tx{chg}(\tau^i),\tx{u}^{i}_\tx{chg}(\tau^i),\tau^i)\leq 0, \quad s\in s^{i}_\tx{chg}\label{eq:phyb_gtaucns}\\
&\tx{x}_\tx{drv}(s)\in\mathcal{X}_\tx{drv}(s),\quad \tx{u}_\tx{drv}(s)\in\mathcal{U}_\tx{drv}(s),\quad s\in\mathcal{S}_\tx{drv}\label{eq:phyb_drv_bcns}\\
&\tx{x}^{i}_\tx{chg}(\tau^i)\in\mathcal{X}^{i}_\tx{chg}(\tau^i),\quad \tx{u}^{i}_\tx{chg}(\tau^i)\in\mathcal{U}^{i}_\tx{chg}(\tau^i),\quad s\in s^{i}_\tx{chg}\label{eq:phyb_chg_bcns}\\
&t_\tx{chg}^i\in[0,t_\tx{chg}^{\max}]\label{eq:maxtchg}\\
&\tx{x}^i_\tx{chg}(0)=\tx{x}_\tx{drv}(s^{i}_\tx{chg})-\zeta_i\tx{x}_\tx{detour}\label{eq:drv2chg}\\
&\tx{x}_\tx{drv}(s_\tx{chg}^{i^{+}})=\tx{x}^{i}_\tx{chg}(1)-\zeta_i\tx{x}_\tx{detour}\label{eq:chg2drv}\\
&\tx{x}_\tx{drv}(s_0)\in\mathcal{X}_\tx{drv0}, \quad \tx{x}_\tx{drv}(s_\tx{f})\in\mathcal{X}_\tx{drvf}
\end{align}
\end{subequations}}%
where $t_\tx{chg}^i$ and $\zeta_i$ are considered as design parameters, $\tx{s}_\tx{0}$ and $\tx{s}_\tx{f}$ denote initial and final vehicle position, respectively, $t_\tx{chg}^{\max}$ is maximum allowed charging time, $g_\tx{drv}$ and $g_\tx{chg}$ represent the battery power limits \eqref{eq:pbbound} during driving and charging modes, respectively, and $s_\tx{chg}^{i^{+}}$ is an instance denoting the vehicle's position when charging is done and the vehicle is leaving the charging station. Also, $\mathcal{X}_\tx{drv}$ and $\mathcal{U}_\tx{drv}$ are the feasible sets of states and control inputs for the driving mode, and $\mathcal{X}_\tx{chg}$ and $\mathcal{U}_\tx{chg}$ are the corresponding feasible sets for the charging mode. Furthermore, $\mathcal{X}_\tx{drv0}$ and $\mathcal{X}_\tx{drvf}$ denote allowed initial and target states, respectively. Moreover, x$_\tx{detour}$ corresponds to the change in battery temperature and SoC during the detour periods. The constraints \eqref{eq:drv2chg} and \eqref{eq:chg2drv} represent the transition between the modes. Accordingly, the battery temperature and SoC at the beginning of the charging event must be equal to the corresponding variables at the arrival of the charging station. Similarly, the battery temperature and SoC when the vehicle resumes its drive after charging must be equal to the corresponding variables at the end of the charging event. The problem \eqref{eq:phyb} is a mixed-integer nonlinear program (MINLP), due to the binary variable $\zeta$ and nonlinear relations in the constraints and cost function.

\begin{figure}[t!]
 \centering
 \includegraphics[width=\linewidth]{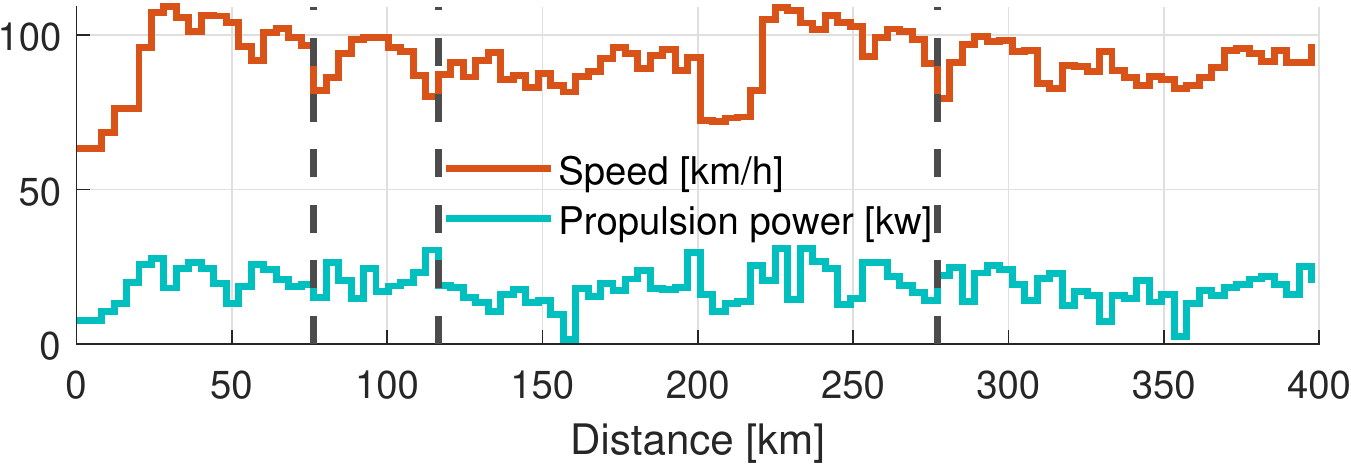}
  \caption{\footnotesize Vehicle drive cycle including the vehicle speed and propulsion power trajectories. Dashed vertical lines indicate available charging locations.}
  \label{fig:dc}
\end{figure}

\section{Results}\label{sec:res}
In this section, simulation results are provided for the BEV demonstrated in Fig.~\ref{fig:scenario}. Within the simulations, we investigate the benefits of including a heat pump in the TM system. Also, we consider the charge point planning, in favour of achieving an optimal compromise between time and energy cost. The simulation setup and the results are given in Section~\ref{subsec:setup} and Sections~\ref{subsec:evst} to \ref{subsec:evstamb}, respectively.

\subsection{Simulation Setup}\label{subsec:setup}
As depicted in Fig.~\ref{fig:dc}, the simulations are conducted on a \SI{400}{km} long drive cycle, which is based on real-world measurements. Three available charging locations along the driving route are marked by dashed vertical lines. The used charging stops are indicated with a solid vertical line hereafter. The vehicle starts its drive with a battery soaked in the ambient temperature, i.e. $T_{\tx{b0}}=T_{\tx{amb}}$. Also, cabin climate and auxiliary load demand are supplied during both the driving and charging modes. Furthermore, the cost for occupying the charging spot is assumed to be zero, i.e. $c_\tx{T}=0$. The results shown in the remainder of the paper use those vehicle and simulation parameters reported in Table \ref{tab:par}, unless stated otherwise.

The MINLP \eqref{eq:phyb} is discretised with a distance sampling interval of \SI{4}{km}, using the Runge-Kutta $4^{th}$ order method~\cite{butcher76}. The discretised problem is solved with the solver BONMIN, using the open source nonlinear optimisation tool CasADi~\cite{andersson19} in Matlab.

\begin{table}
\begin{center}
\caption{Vehicle and Simulation Parameters} 
\label{tab:par}
\begin{tabular}{l l}
\hline
Maximum battery capacity & $C_\tx{b}=\SI{195}{Ah}$\vspace{0.05cm}\\ 
Product of specific heat and battery mass & $c_\tx{p}m_\tx{b}=\SI{375}{kJ/(K)}$\vspace{0.05cm}\\ 
Route length & $\SI{400}{km}$\vspace{0.05cm} 
\\
Distance sampling interval & $\SI{4}{km}$\vspace{0.05cm} 
\\
Number of charging along the route  & $N_\tx{chg}=\SI{3}{}$\vspace{0.05cm} \\
Detour time for each charging stop  & $t_\tx{d}=\SI{300}{s}$\vspace{0.05cm} \\
Detour energy for each charging stop  & $E_\tx{d}=\SI{450}{Wh}$\vspace{0.05cm} \\
Electrical energy cost while charging & $c_\tx{e}=\SI{8.7}{SEK/kWh}$\vspace{0.05cm} \\
Charger rated power & $P^{\max}_\tx{grid}=\SI{200}{kW}$\vspace{0.05cm} \\
Auxiliary load & $P_\tx{aux}=\SI{0.5}{kW}$\vspace{0.05cm} \\
% HVCH power for heating cabin & $P^\tx{c}_{\tx{hvch}}=\SI{1.5}{kW}$ \vspace{0.05cm}\\
Maximum HVCH power & $P^{\max}_{\tx{hvch}}=\SI{7}{kW}$\vspace{0.05cm} \\
Maximum HVAC power & $P^{\max}_{\tx{hvac}}=\SI{3}{kW}$ \vspace{0.05cm}\\
Maximum HP power & $P^{\max}_{\tx{hp}}=\{0,1,3\} \tx{kW}$ \vspace{0.05cm}\\
HVCH power to heat rate efficiency & $\eta_{\tx{hvch}}=\SI{87}{\%}$\vspace{0.05cm} \\
%HVAC power to heat rate efficiency & $\eta_{\tx{hvac}}=\SI{87}{\%}$\vspace{0.05cm} \\
Ambient temperature & $T_\tx{amb}=\{-10,0,10\} {^\circ \tx{C}}$\vspace{0.05cm} \\
Initial battery temperature & $T_{\tx{b0}}=T_\tx{amb}$\vspace{0.05cm} \\
Initial battery state of charge & $\tx{soc}_0=\SI{90}{\%}$\vspace{0.05cm} \\
Terminal battery state of charge & $\tx{soc}_\tx{f}=\SI{10}{\%}$\vspace{0.05cm} \\
\hline
\end{tabular}
\end{center}
\end{table}

\begin{figure}[t!]
\centering
\subfigure[Ambient temperature of \SI{-10}{^\circ C}.]{
 \includegraphics[width=.9\linewidth]{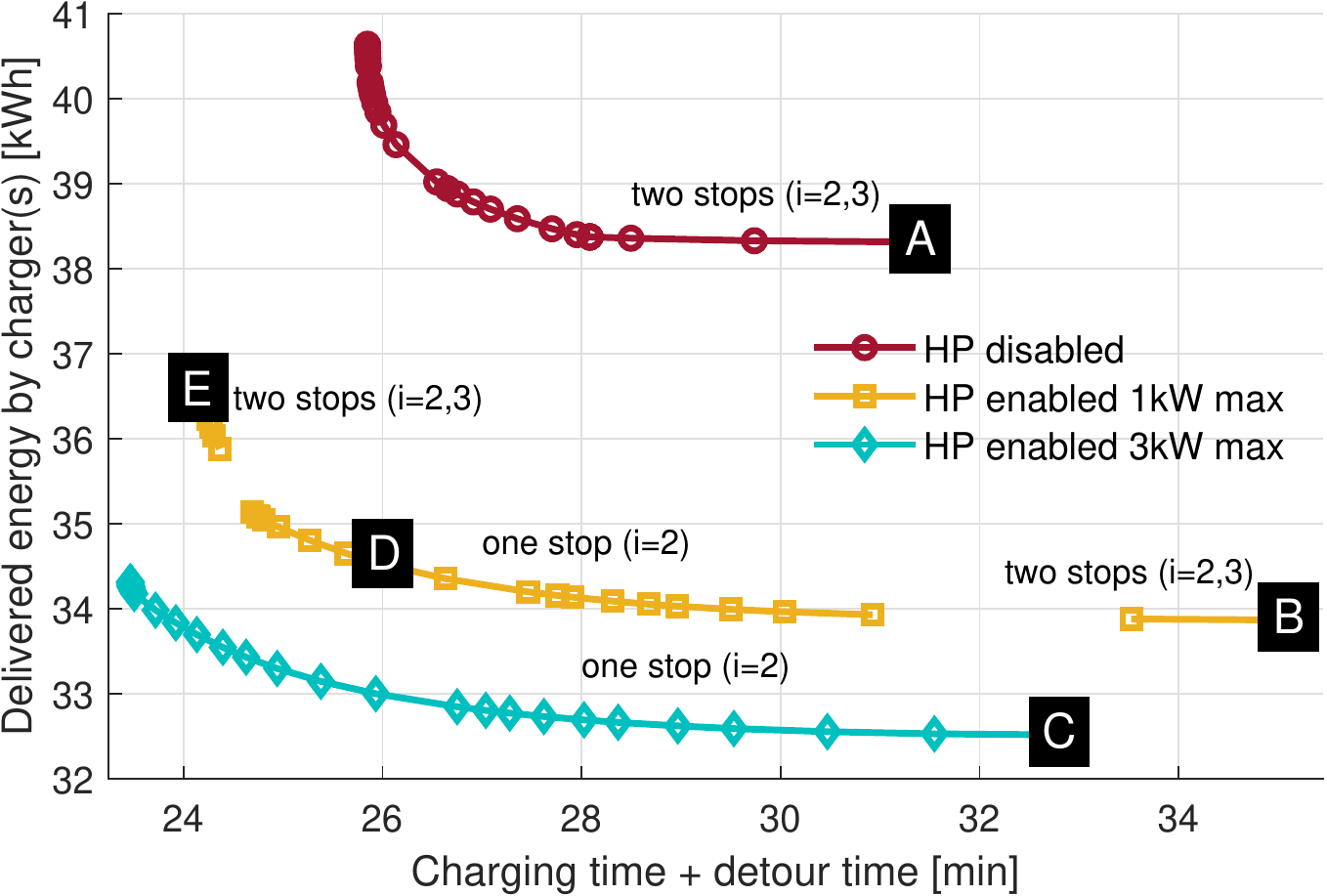}
\label{fig:pf-10}
}
\subfigure[Ambient temperature of \SI{0}{^\circ C}.]{
 \includegraphics[width=.9\linewidth]{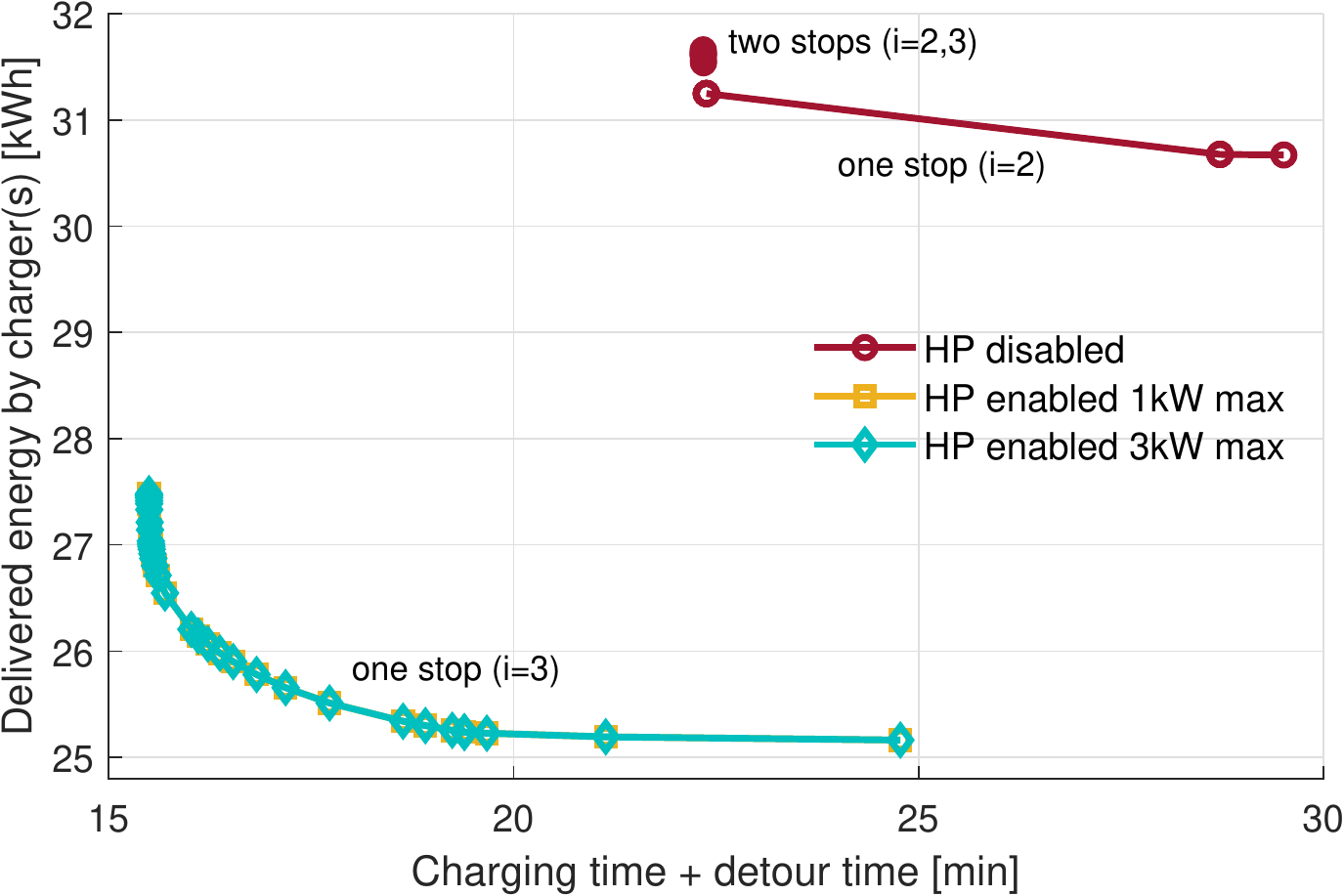}
\label{fig:pf0}
}
\subfigure[Ambient temperature of \SI{10}{^\circ C}.]{

 \includegraphics[width=.9\linewidth]{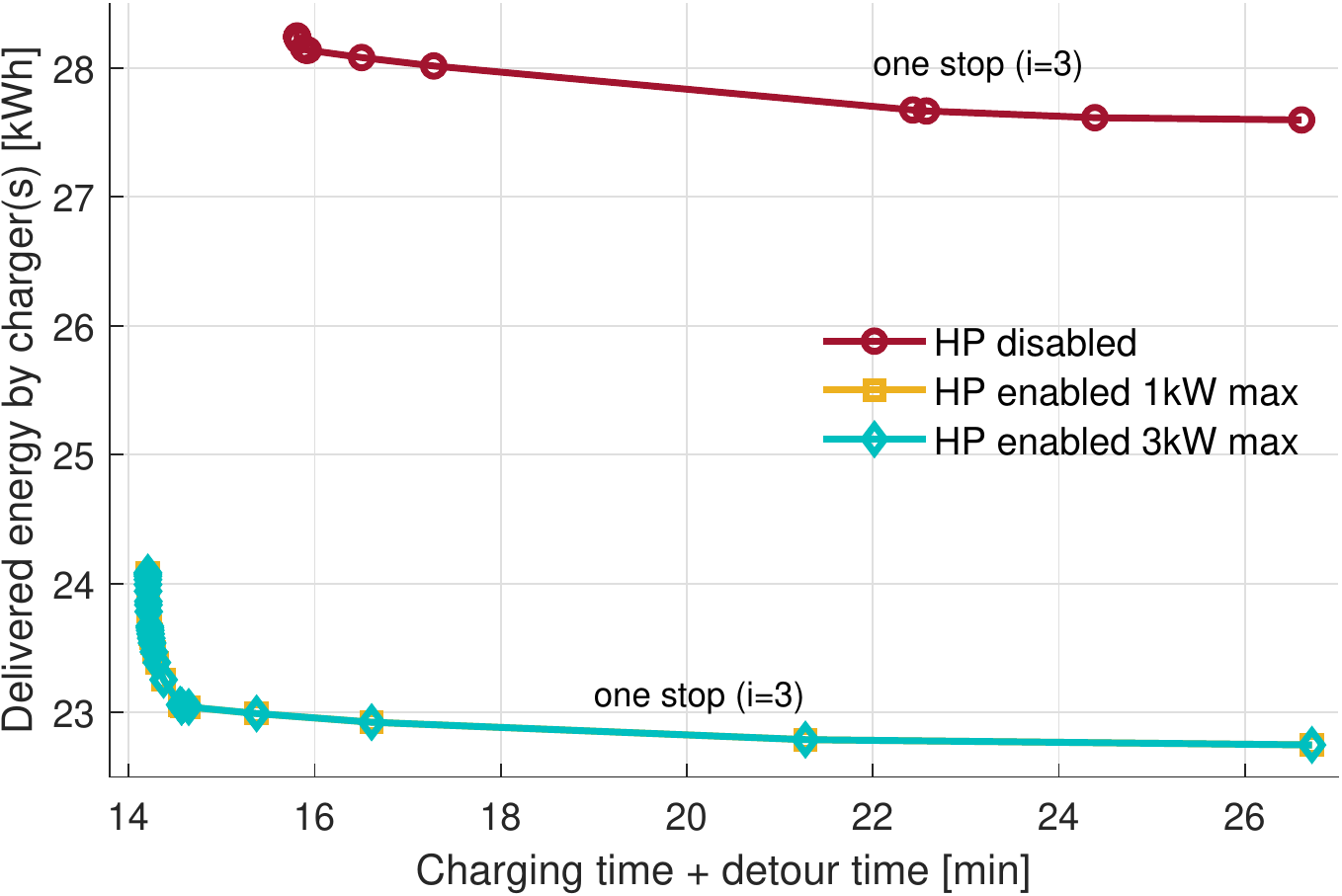}
\label{fig:pf10}
}
\caption{Pareto frontier describing the trade-off between total charging energy versus time including charging and detour times for various ambient temperatures and heat pump power limits.}
\label{fig:pf}
\end{figure}

\subsection{Time vs. Energy Efficiency}\label{subsec:evst}
For different HP power limits and ambient temperatures, the Pareto frontiers are derived describing the trade-off between total charged energy versus combined charging and detour time, as depicted in Fig.~\ref{fig:pf}. The HP is either disabled, or activated with maximum power of \SI{1}{kW} or \SI{3}{kW}, hereafter referred to as {\it{smaller}} HP or {\it{larger}} HP, respectively. To obtain the Pareto graphs, the time cost $c_\tx{t,chg}$ is varied over a large span to obtain solutions that vary respectively from energy optimal to time optimal.

\begin{table}[t]
\caption{Energy reduced solution at different ambient temperatures} \label{tab:eopttab}
  \centering
  \setlength\tabcolsep{6pt}
 \begin{tabular}{|c|c @{\hspace{-.5ex}} c @{\hspace{-.5ex}} c|} 
 \hline
 \multicolumn{4}{| c |}{\textbf{-10 $^\circ$C ambient temperature}}\\
 \hline
 \begin{tabular}{c}Variable\end{tabular} & \begin{tabular}{c}HP disabled\end{tabular}& \begin{tabular}{c}smaller HP\end{tabular}&
 \begin{tabular}{c}larger HP\end{tabular}\\
 \hline
Energy (at \SI{28}{min}) [kWh]& 38.4 & 34.1 & 32.7 \\
Reduction [\%] & - & 11.1 & 14.9 \\
\hline
\hline
 \multicolumn{4}{| c |}{\textbf{0 $^\circ$C ambient temperature}}\\
 \hline
 \begin{tabular}{c}Variable\end{tabular} & \begin{tabular}{c}HP disabled\end{tabular}& \begin{tabular}{c}smaller HP\end{tabular}&
 \begin{tabular}{c}larger HP\end{tabular}\\
 \hline
Energy (at \SI{22}{min}) [kWh]& 31.3 & 25.2 & 25.2 \\
Reduction [\%] & - & 19.4 & 19.4 \\
\hline
\hline
 \multicolumn{4}{| c |}{\textbf{10 $^\circ$C ambient temperature}}\\
 \hline
 \begin{tabular}{c}Variable\end{tabular} & \begin{tabular}{c}HP disabled\end{tabular}& \begin{tabular}{c}smaller HP\end{tabular}&
 \begin{tabular}{c}larger HP\end{tabular}\\
 \hline
Energy (at \SI{16}{min}) [kWh]& 28.0 & 22.3 & 22.3 \\
Reduction [\%] & - & 18.4 & 18.4 \\
\hline
\end{tabular}
\end{table}

\subsection{Energy Optimal Trip}\label{subsec:eopt}
From the Pareto frontiers shown in Fig.~\ref{fig:pf-10}-Fig.~\ref{fig:pf10}, it is observable that activating HP generally leads to reduced energy consumption. For instance, at \SI{-10}{^\circ C} ambient temperature and \SI{28}{min} of combined charging and detour time, the charger(s) delivered energy is decreased by \SI{11}{\%} for the \SI{1}{kW} limited HP and \SI{15}{\%} for the larger HP, compared to the similar scenario but with the HP disabled. Such energy reduction is due to the HP being used to move the heat from the battery loop into the cabin compartment, thus reducing the need for the HVCH to be used for cabin heating. The detailed results of charged energy together with the energy reduction percentage for different ambient temperatures and HP power limits are given in Table \ref{tab:eopttab}. Furthermore, at a given ambient temperature, the number of charging stops may change for different HP maximum power values. Also, it is observed that a more powerful HP is more beneficial compared to the smaller HP, at low ambient temperatures. However, at high ambient temperature there is no noticeable advantage of using the larger HP rather than the smaller HP. This will be discussed in more details later in Section~\ref{subsec:evstamb}. 

According to Fig.~\ref{fig:pf-10}, we look more closely at the energy optimal cases at \SI{-10}{^\circ C} ambient temperature, as: 
\begin{itemize}
    \item Case A: energy optimal solution with HP disabled
    \item Case B: energy optimal solution with \SI{1}{kW} HP power limit
    \item Case C: energy optimal solution with \SI{3}{kW} HP power limit
\end{itemize}

\subsubsection{Case A}\label{subsec:case A}
\begin{figure*}[t!]
\centering
\subfigure[Battery temperature and SoC trajectories vs. travelled distance.]{
 \includegraphics[width=.45\linewidth]{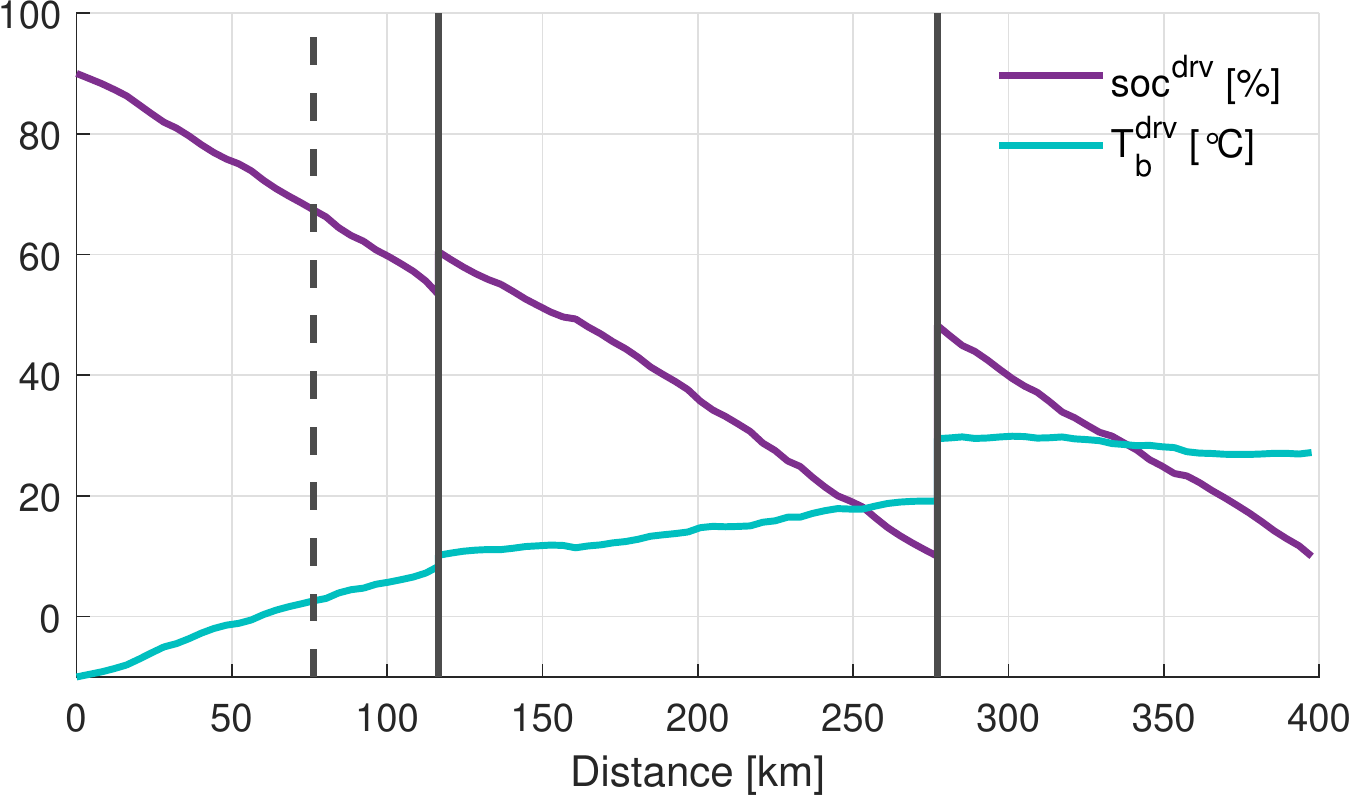}\hspace{.25cm}
\label{fig:caseA_soctbdrv}
}
\subfigure[Trajectories of battery power and propulsion power together with battery power limits vs. travelled distance.]{

 \includegraphics[width=.47\linewidth]{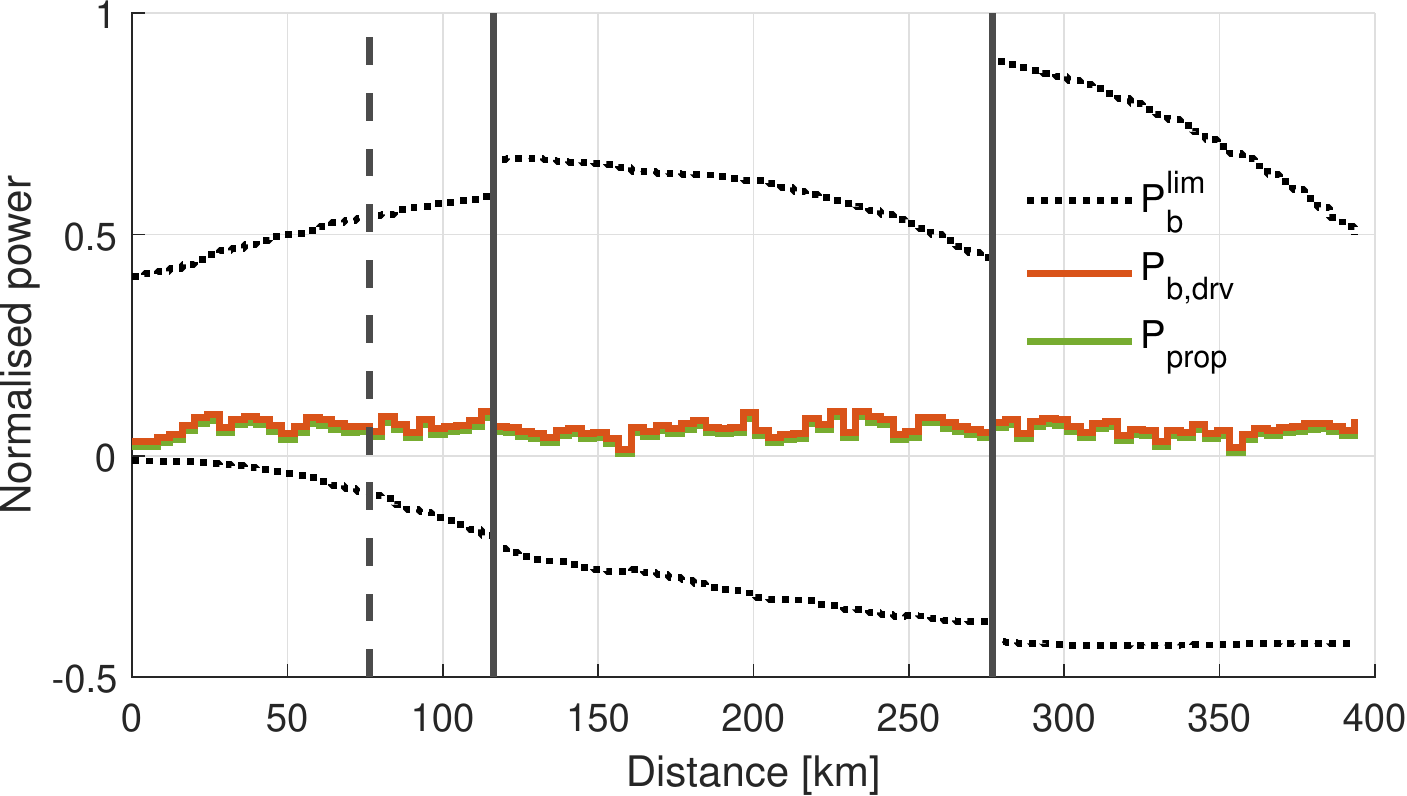}
\label{fig:caseA_pbpropdrv}
}
\subfigure[Trajectories of HVCH and HP power for cabin heating, vs. travelled distance.]{

 \includegraphics[width=.47\linewidth]{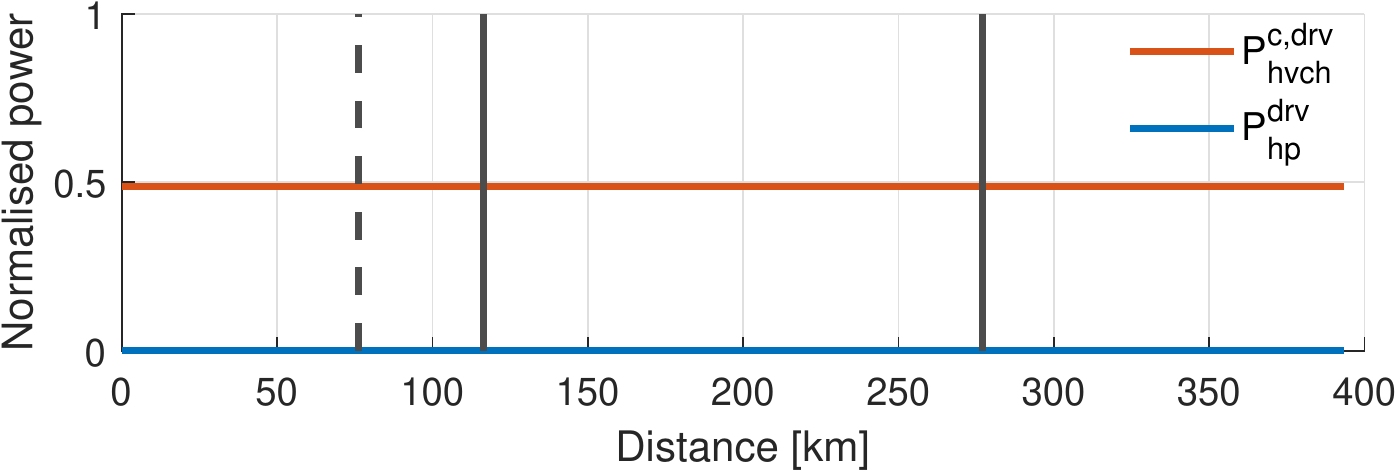}
\label{fig:caseA_hvchchpdrv}
}
\subfigure[Trajectories of HVCH and HVAC power, respectively for battery heating and cooling, vs. travelled distance.]{

 \includegraphics[width=.47\linewidth]{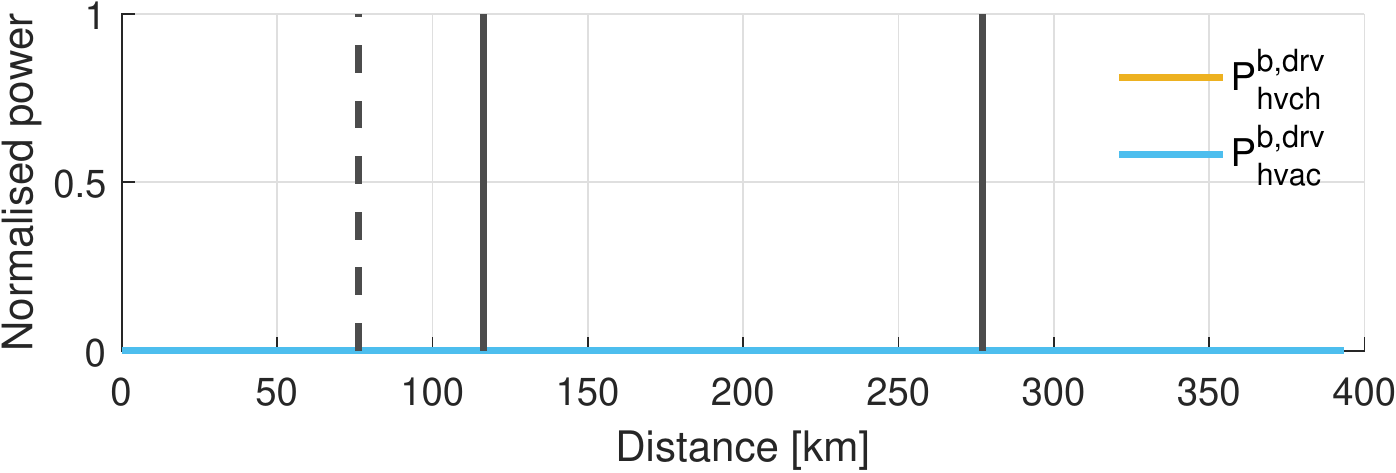}
\label{fig:caseA_hvchbhvacbdrv}
}
\subfigure[Battery temperature and SoC trajectories vs. charging time (second charging stop).]{

 \includegraphics[width=.47\linewidth]{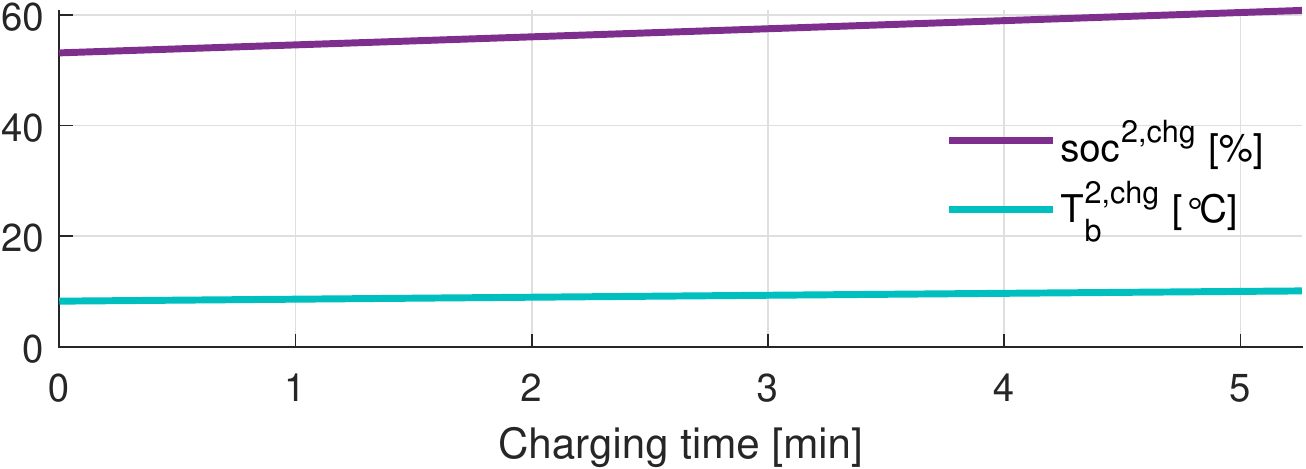}
\label{fig:caseA_soctbchg2}
}
\subfigure[Trajectories of HVCH and HP power for cabin heating, vs. charging time (second charging stop).]{

 \includegraphics[width=.47\linewidth]{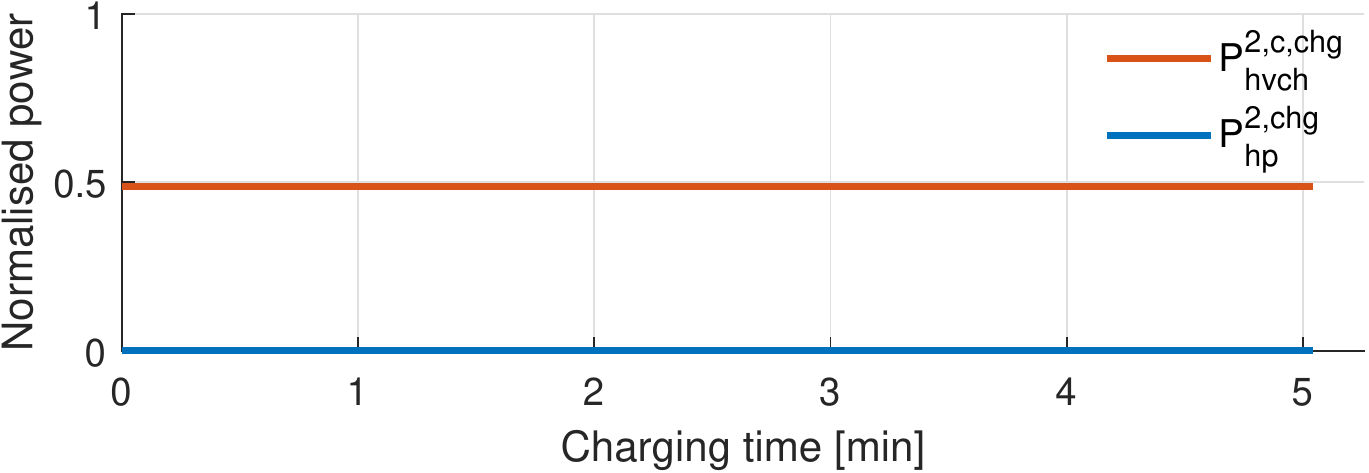}
\label{fig:caseA_hvchchpchg2}
}
\subfigure[Trajectories of HVCH and HVAC power, respectively for battery heating and cooling, vs. charging time (second charging stop).]{

 \includegraphics[width=.47\linewidth]{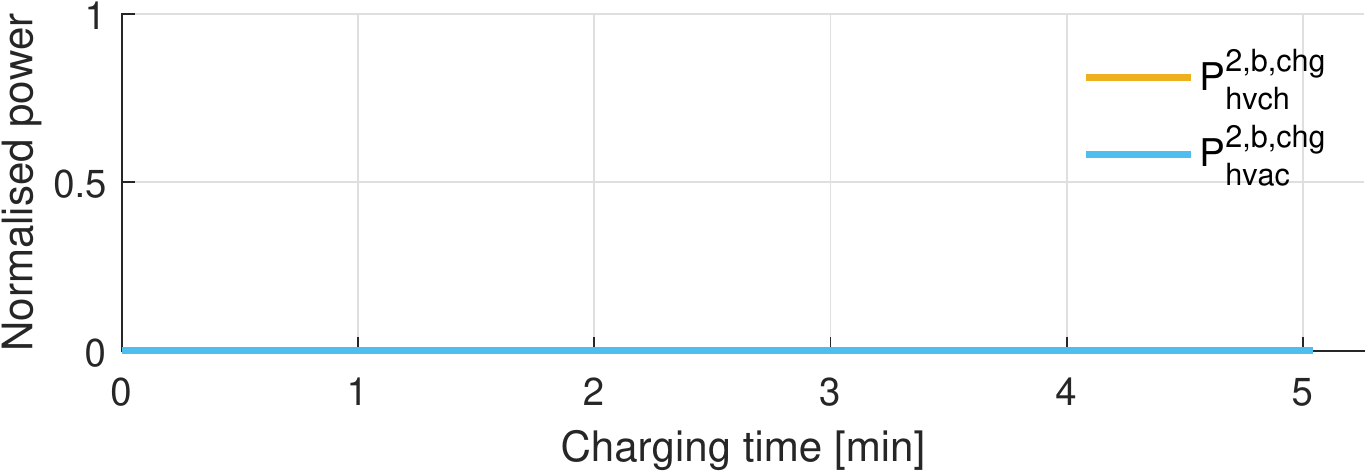}
\label{fig:caseA_hvchbhvacbchg2}
}
\subfigure[Battery temperature and SoC trajectories vs. charging time (third charging stop).]{

 \includegraphics[width=.47\linewidth]{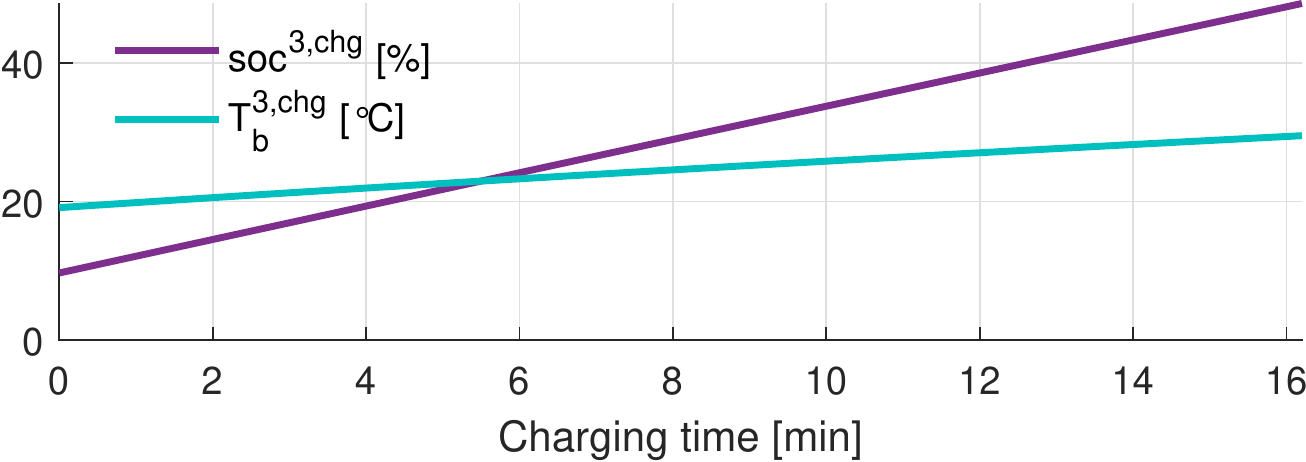}
\label{fig:caseA_soctbchg3}
}
\subfigure[Trajectories of HVCH and HP power for cabin heating, vs. charging time (third charging stop).]{

 \includegraphics[width=.47\linewidth]{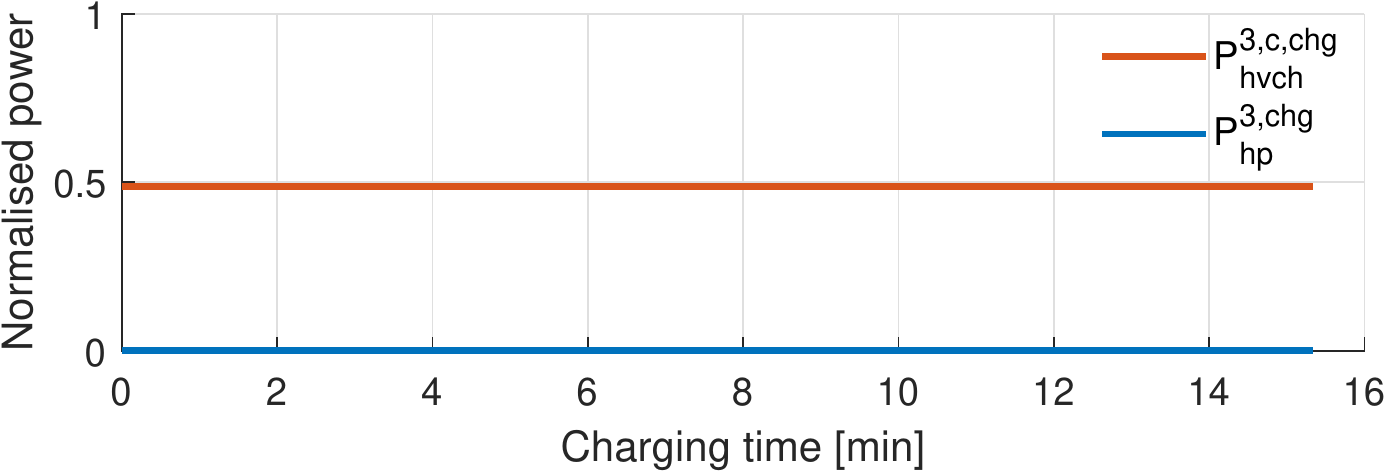}
\label{fig:caseA_hvchchpchg3}
}
\subfigure[Trajectories of HVCH and HVAC power, respectively for battery heating and cooling, vs. charging time (third charging stop).]{

 \includegraphics[width=.47\linewidth]{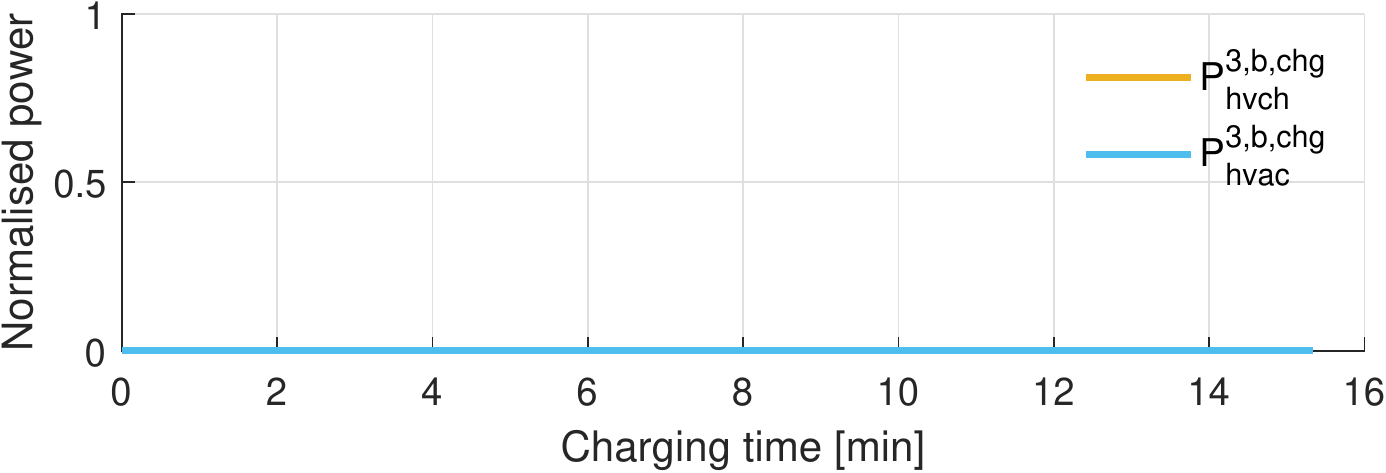}
\label{fig:caseA_hvchbhvacbchg3}
}
\caption{Case A: Energy optimal case with heat pump disabled.}
\label{fig:caseA}
\end{figure*}

States and control inputs trajectories versus travelled distance and charging time are depicted in Fig.~\ref{fig:caseA}, where the power is normalised with the maximum HVCH power. In this case, it is optimal to select two charging occasions ($i=2,3$) along the trip. The battery temperature increases significantly over the course of the trip and levels out between $\SI{25}{^\circ C}$ and $\SI{30}{^\circ C}$ at the destination, whereas no active battery heating is done with the HVCH. Such battery temperature increase is only due to the passively generated heat, which is mainly kept within the battery pack, and not pumped to the cabin by HP. Active cooling by HVAC is not used in this case, since the battery is kept below the maximum allowed temperature of $\SI{40}{^\circ C}$, by just exchanging the heat to the ambient air. The battery discharge power limit is kept at reasonable levels, as shown in Fig.~\ref{fig:caseA_pbpropdrv}, which is due to the overall high battery temperature throughout the trip.

\subsubsection{Case B}\label{subsec:case B}
\begin{figure*}[t!]
\centering
\subfigure[Battery temperature and SoC trajectories vs. travelled distance.]{
 \includegraphics[width=.45\linewidth]{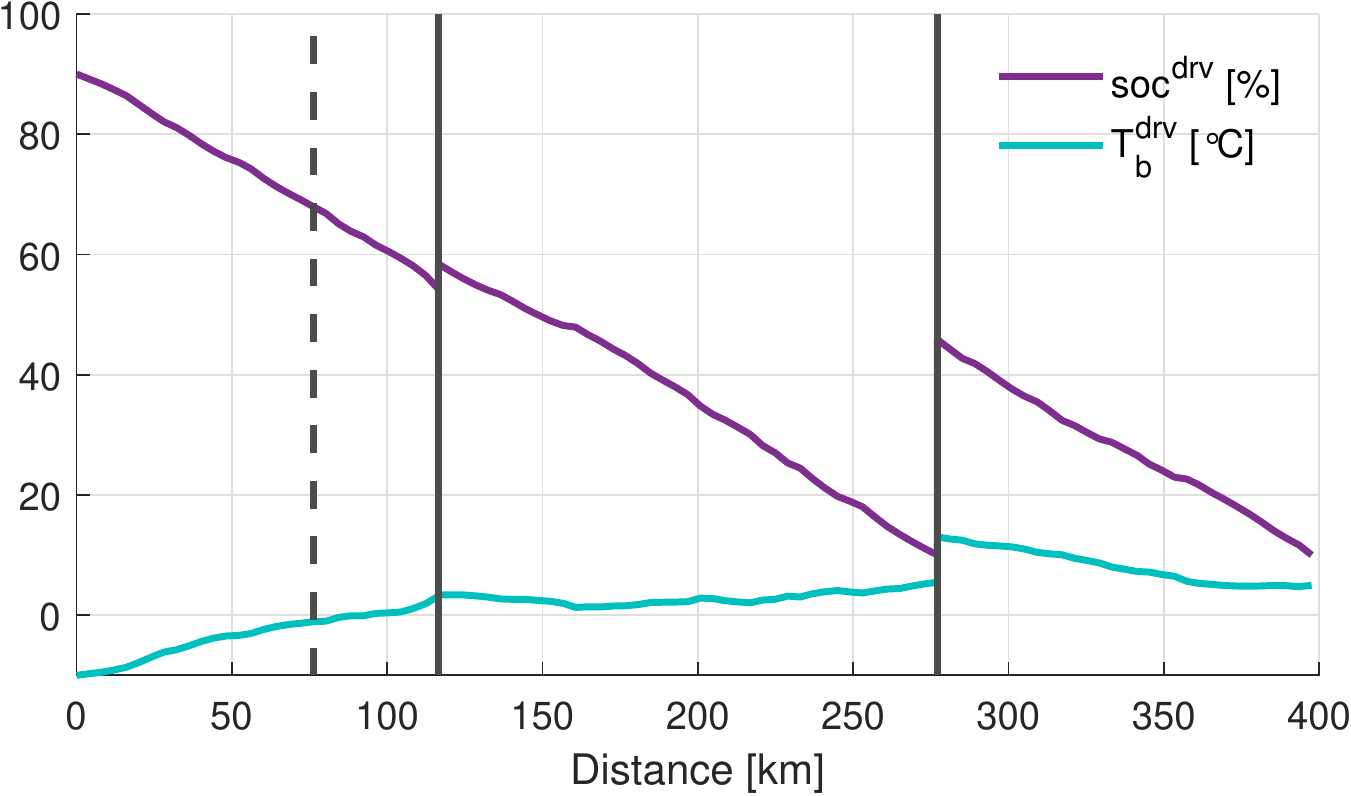}\hspace{.25cm}
\label{fig:caseB_soctbdrv}
}
\subfigure[Trajectories of battery power and propulsion power together with battery power limits vs. travelled distance.]{

 \includegraphics[width=.47\linewidth]{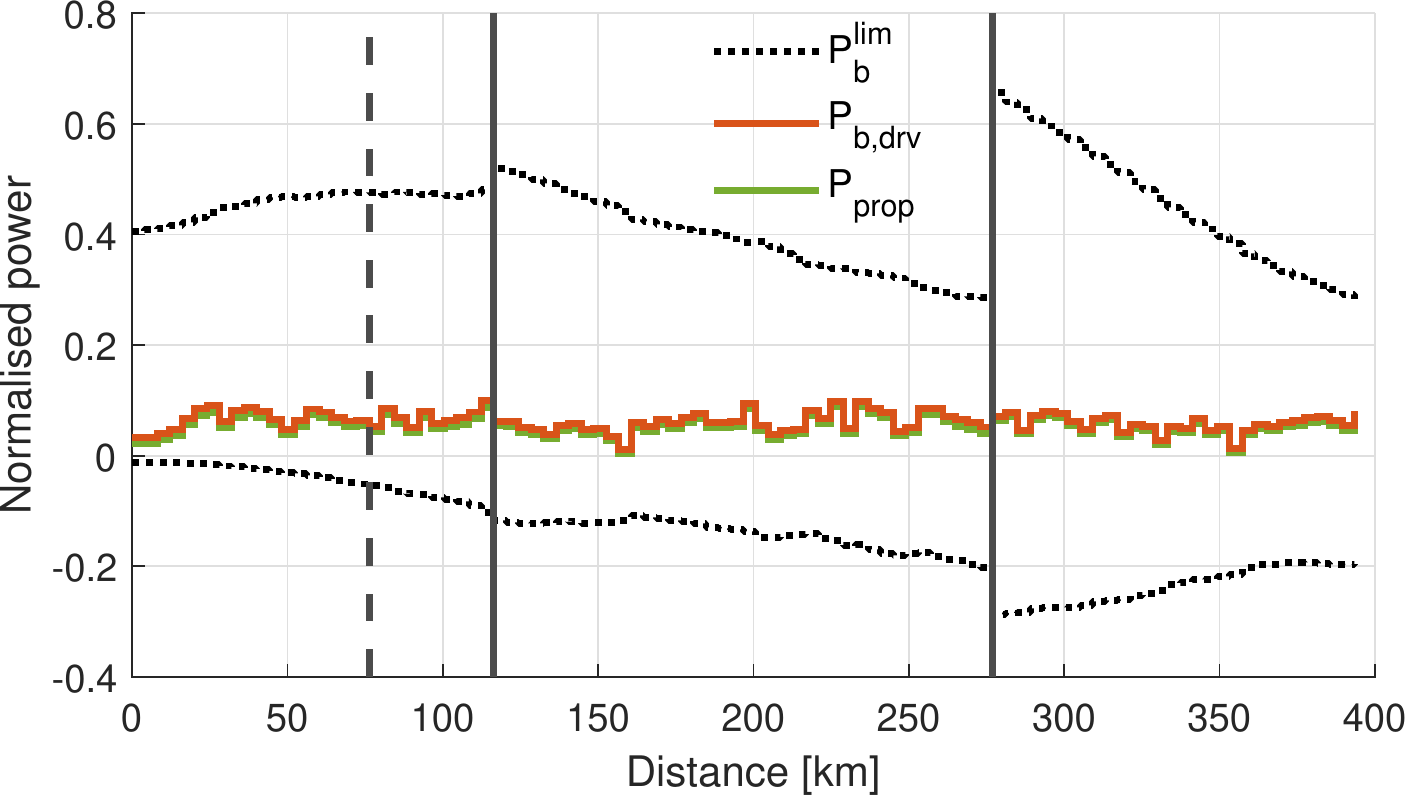}
\label{fig:caseB_pbpropdrv}
}
\subfigure[Trajectories of HVCH and HP power for cabin heating, vs. travelled distance.]{

 \includegraphics[width=.47\linewidth]{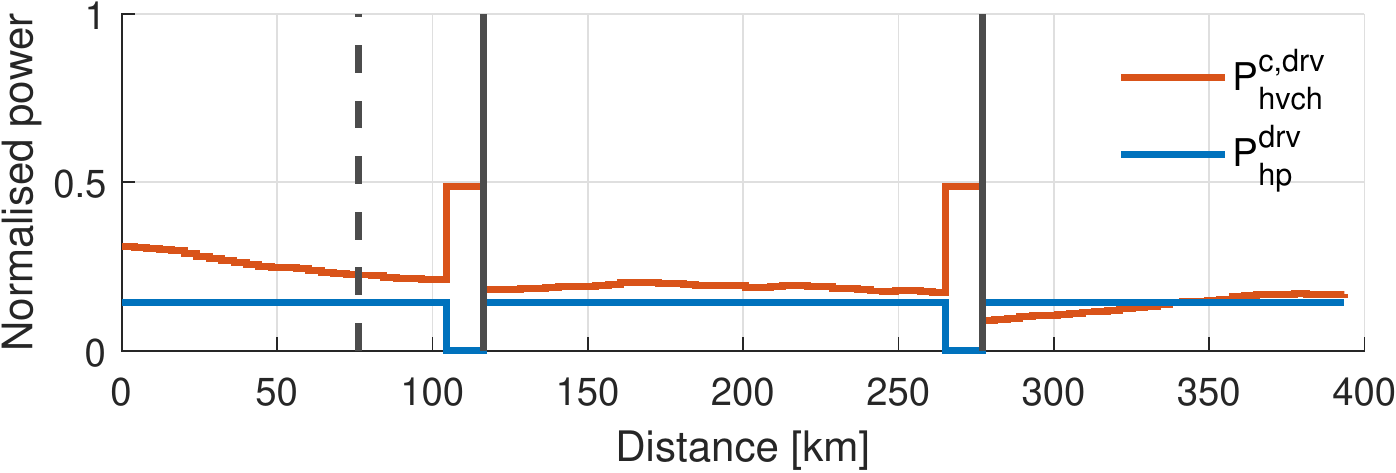}
\label{fig:caseB_hvchchpdrv}
}
\subfigure[Trajectories of HVCH and HVAC power, respectively for battery heating and cooling, vs. travelled distance.]{

 \includegraphics[width=.47\linewidth]{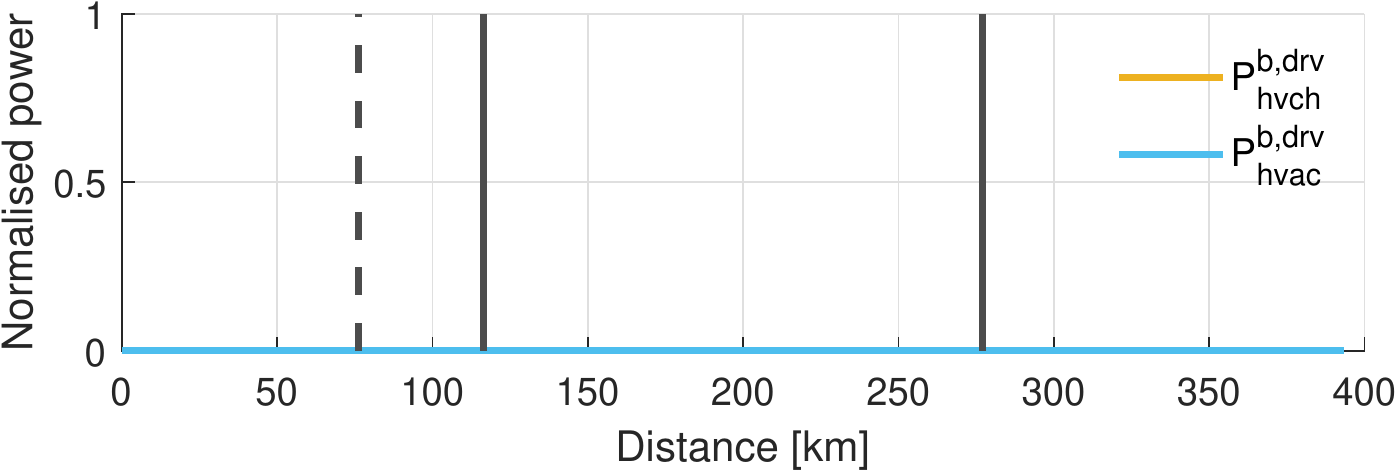}
\label{fig:caseB_hvchbhvacbdrv}
}
\subfigure[Battery temperature and SoC trajectories vs. charging time (second charging stop).]{

 \includegraphics[width=.47\linewidth]{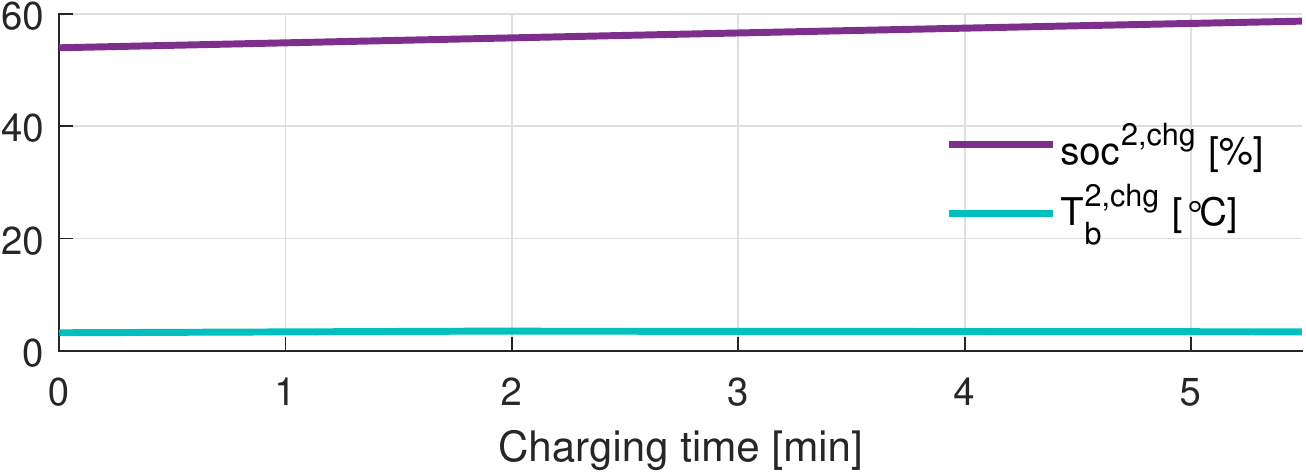}
\label{fig:caseB_soctbchg2}
}
\subfigure[Trajectories of HVCH and HP power for cabin heating, vs. charging time (second charging stop).]{

 \includegraphics[width=.47\linewidth]{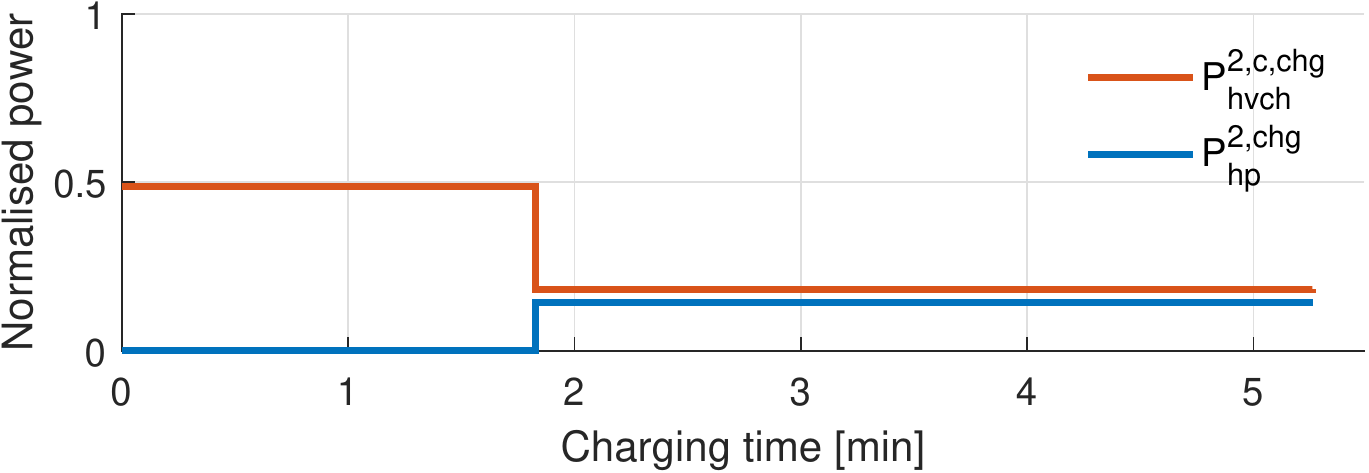}
\label{fig:caseB_hvchchpchg2}
}
\subfigure[Trajectories of HVCH and HVAC power, respectively for battery heating and cooling, vs. charging time (second charging stop).]{

 \includegraphics[width=.47\linewidth]{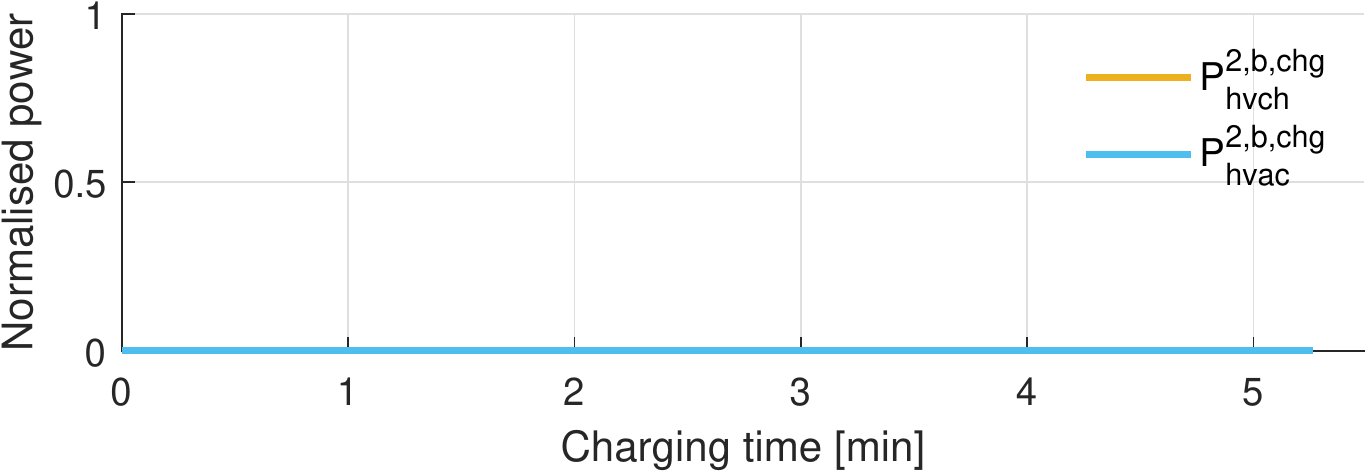}
\label{fig:caseB_hvchbhvacbchg2}
}
\subfigure[Battery temperature and SoC trajectories vs. charging time (third charging stop).]{

 \includegraphics[width=.47\linewidth]{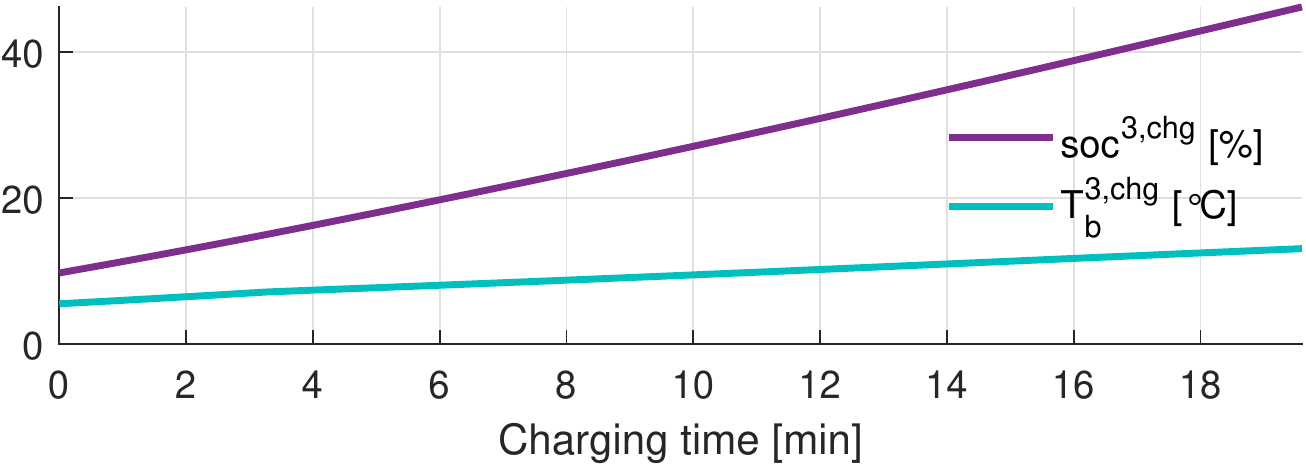}
\label{fig:caseB_soctbchg3}
}
\subfigure[Trajectories of HVCH and HP power for cabin heating, vs. charging time (third charging stop).]{

 \includegraphics[width=.47\linewidth]{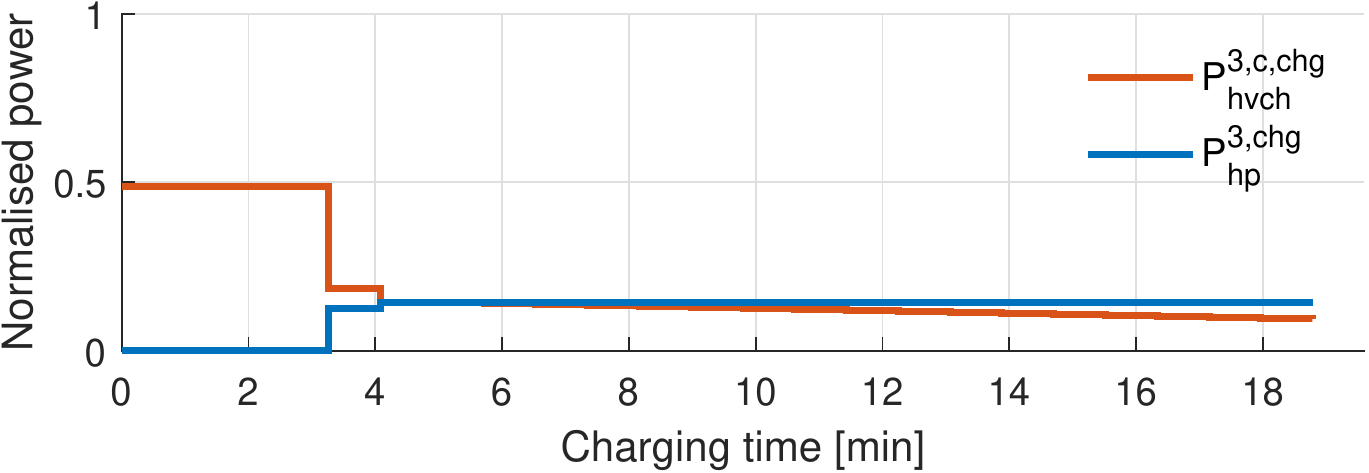}
\label{fig:caseB_hvchchpchg3}
}
\subfigure[Trajectories of HVCH and HVAC power, respectively for battery heating and cooling, vs. charging time (third charging stop).]{

 \includegraphics[width=.47\linewidth]{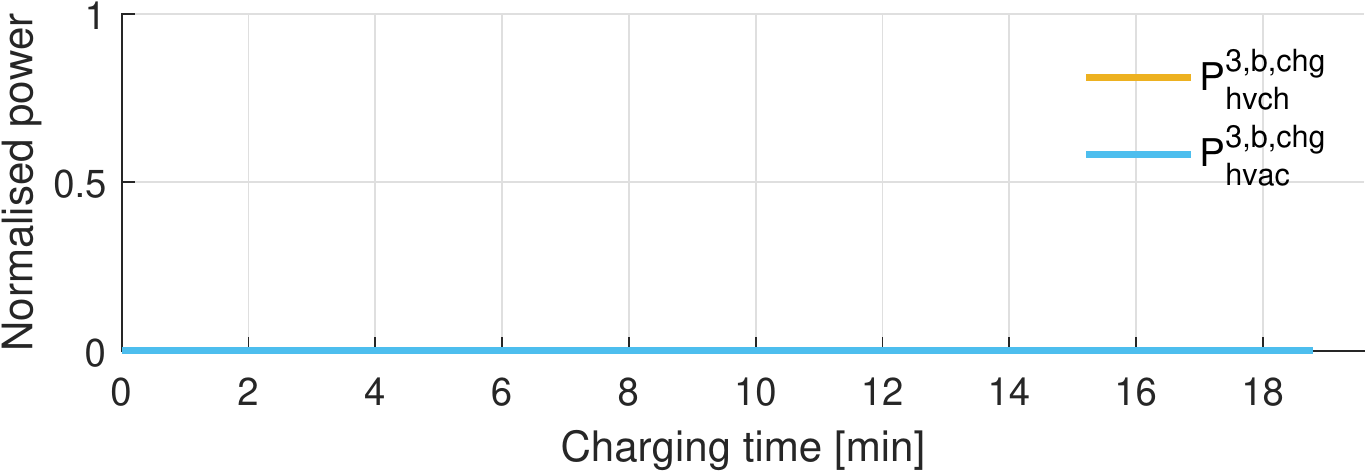}
\label{fig:caseB_hvchbhvacbchg3}
}
\caption{Case B: Energy optimal case with \SI{1}{kW} heat pump power limit.}
\label{fig:caseB}
\end{figure*}

States and control inputs trajectories versus travelled distance and charging time are shown in Fig.~\ref{fig:caseB}. The solution for Case B also involves charging twice ($i=2,3$). This implies that the cost associated with the detour of stopping twice is less than the cost of stopping once at charger $i=2$. Performing only a single charging stop would in this case mean charging in a high SoC region with reduced charging speed at the second charging location. This leads to a longer charging time and more energy spent on maintaining cabin climate and supplying auxiliary load. The HP is switched off right before each charging stop and stays off during a portion of the charging period, as demonstrated in Fig.~\ref{fig:caseB_hvchchpdrv}, Fig.~\ref{fig:caseB_hvchchpchg2}, and Fig.~\ref{fig:caseB_hvchchpchg3}. This means that the Joule and ED losses are prioritised for battery heating right before and at the beginning of the charging periods. The HVCH is not used at all for battery heating in this case in order to minimise unnecessary heat losses to the ambient environment.

\subsubsection{Case C}\label{subsec:case C}
\begin{figure*}[t!]
\centering
\subfigure[Battery temperature and SoC trajectories vs. travelled distance.]{
 \includegraphics[width=.45\linewidth]{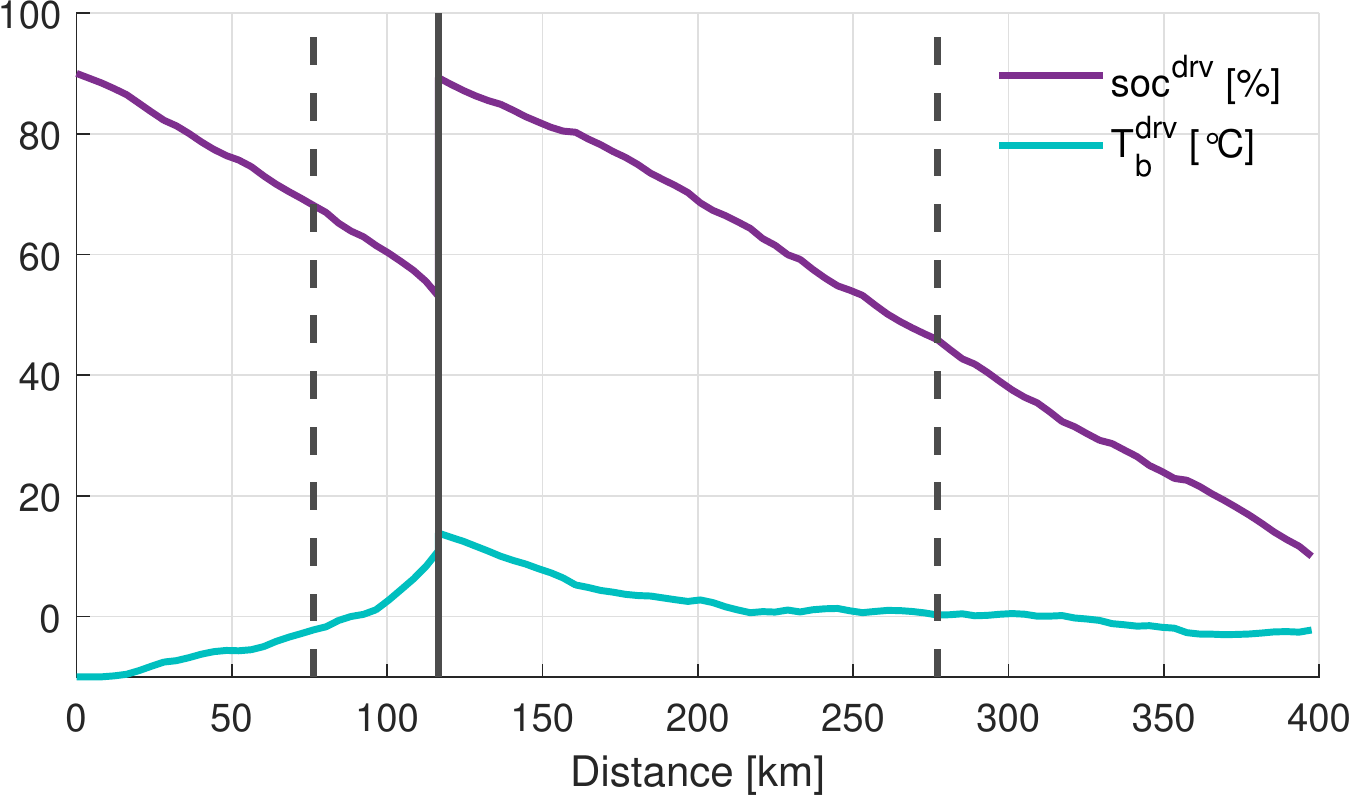}\hspace{.25cm}
\label{fig:caseC_soctbdrv}
}
\subfigure[Trajectories of battery power and propulsion power together with battery power limits vs. travelled distance.]{

 \includegraphics[width=.47\linewidth]{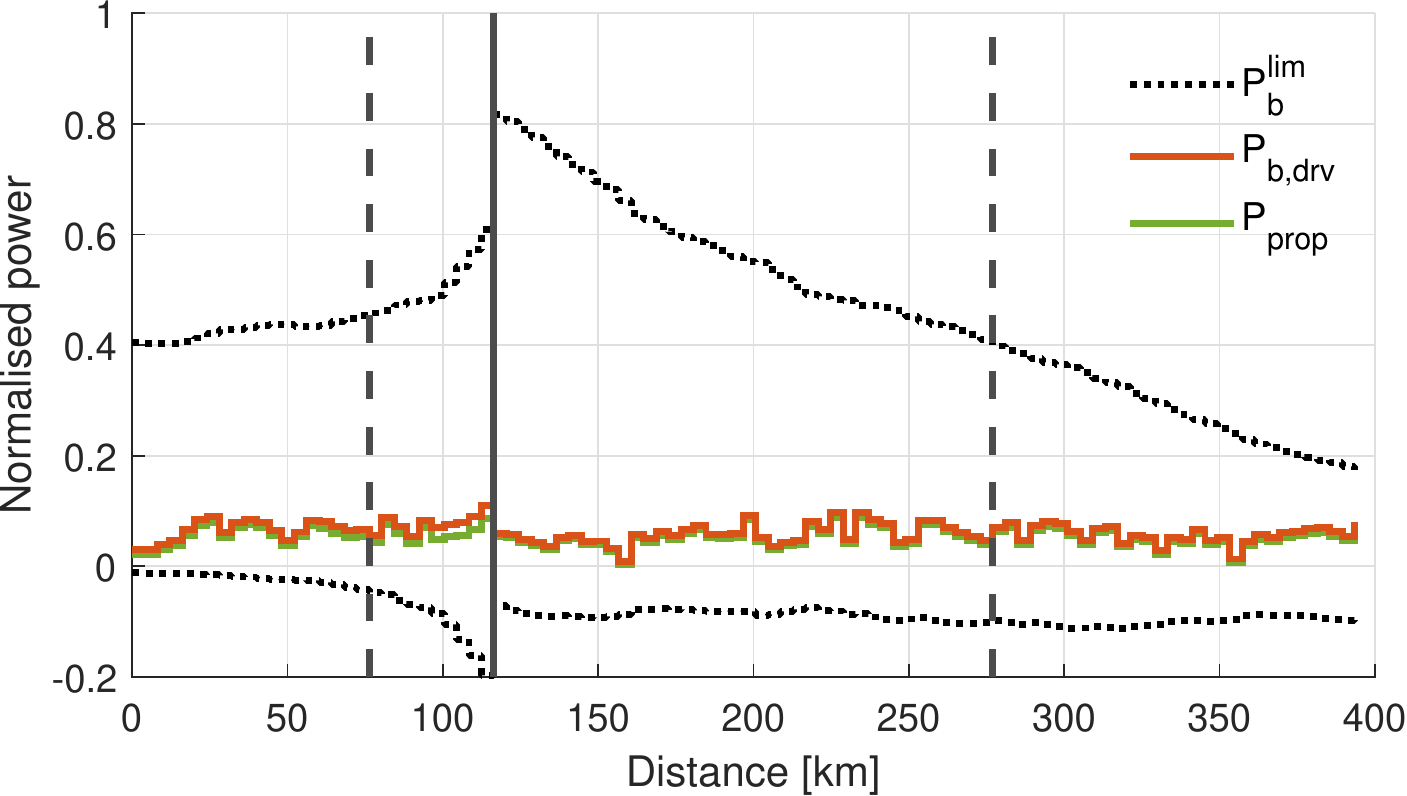}
\label{fig:caseC_pbpropdrv}
}
\subfigure[Trajectories of HVCH and HP power for cabin heating, vs. travelled distance.]{

 \includegraphics[width=.47\linewidth]{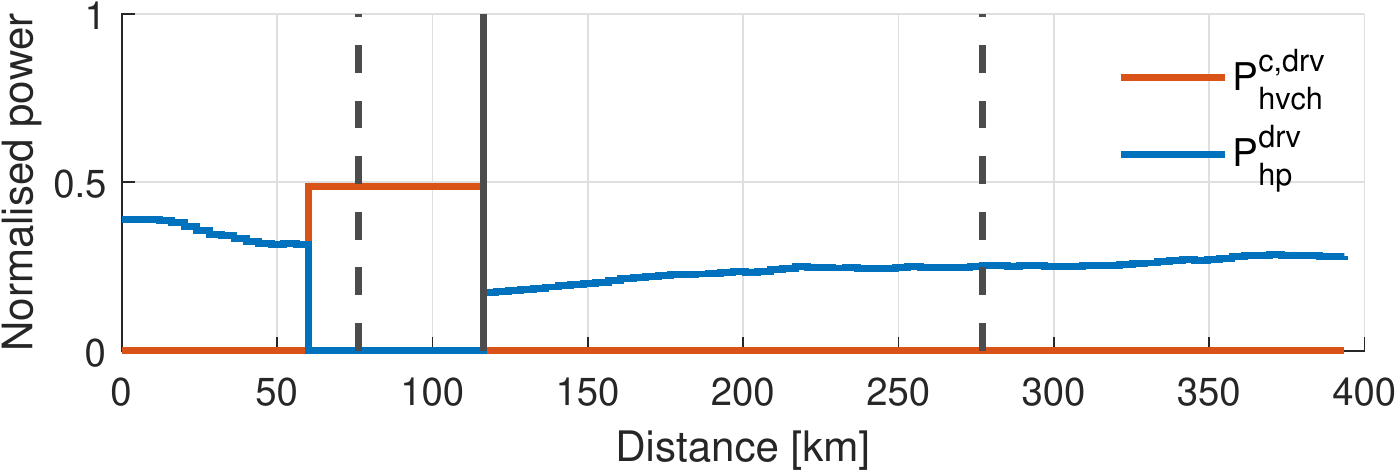}
\label{fig:caseC_hvchchpdrv}
}
\subfigure[Trajectories of HVCH and HVAC power, respectively for battery heating and cooling, vs. travelled distance.]{

 \includegraphics[width=.47\linewidth]{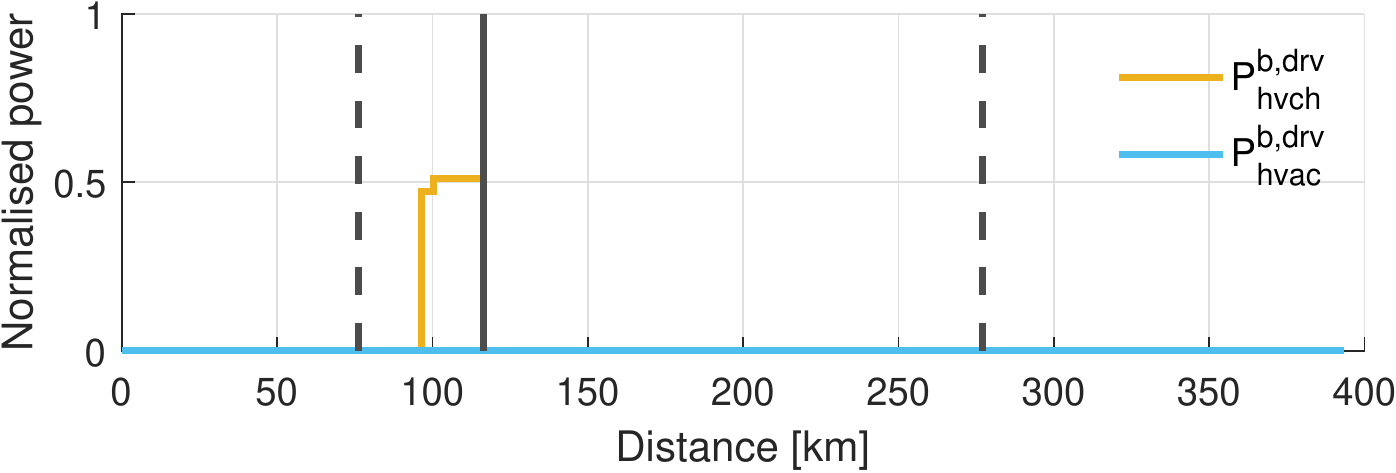}
\label{fig:caseC_hvchbhvacbdrv}
}
\subfigure[Battery temperature and SoC trajectories vs. charging time (second charging stop).]{

 \includegraphics[width=.47\linewidth]{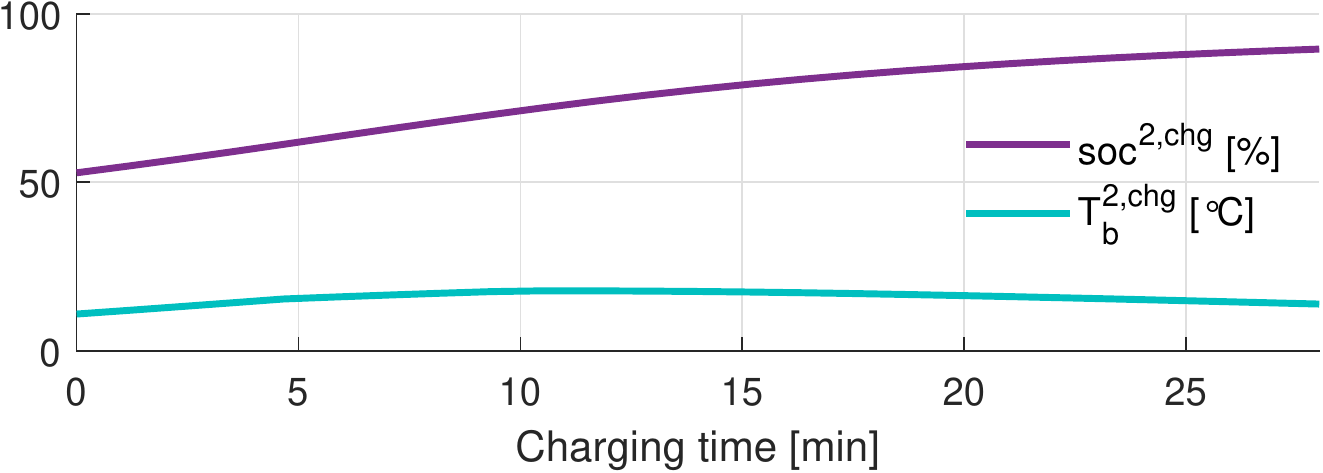}
\label{fig:caseC_soctbchg2}
}
\subfigure[Trajectories of HVCH and HP power for cabin heating, vs. charging time (second charging stop).]{

 \includegraphics[width=.47\linewidth]{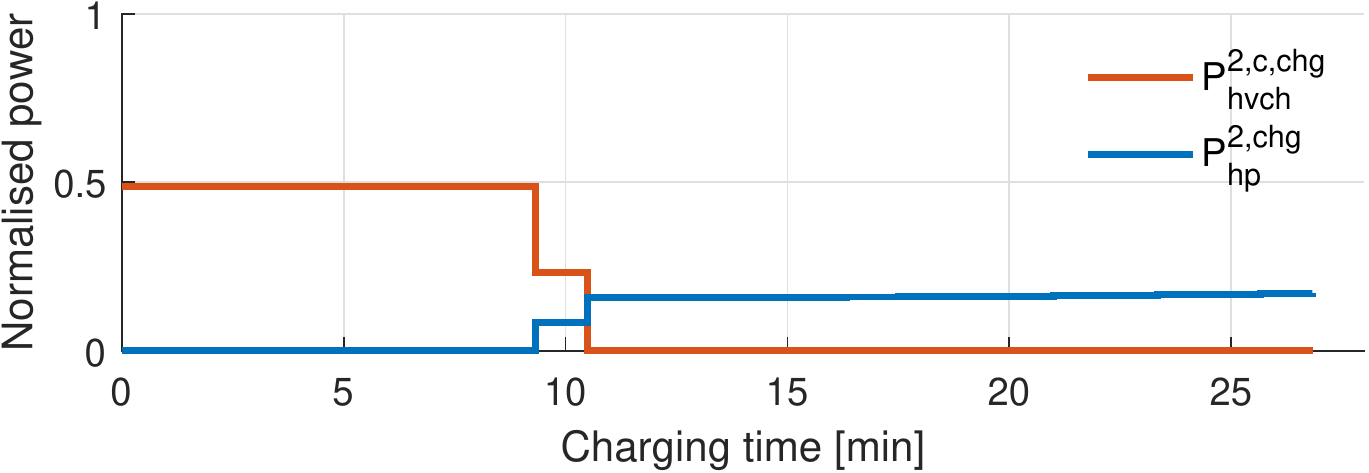}
\label{fig:caseC_hvchchpchg2}
}
\subfigure[Trajectories of HVCH and HVAC power, respectively for battery heating and cooling, vs. charging time (second charging stop).]{

 \includegraphics[width=.47\linewidth]{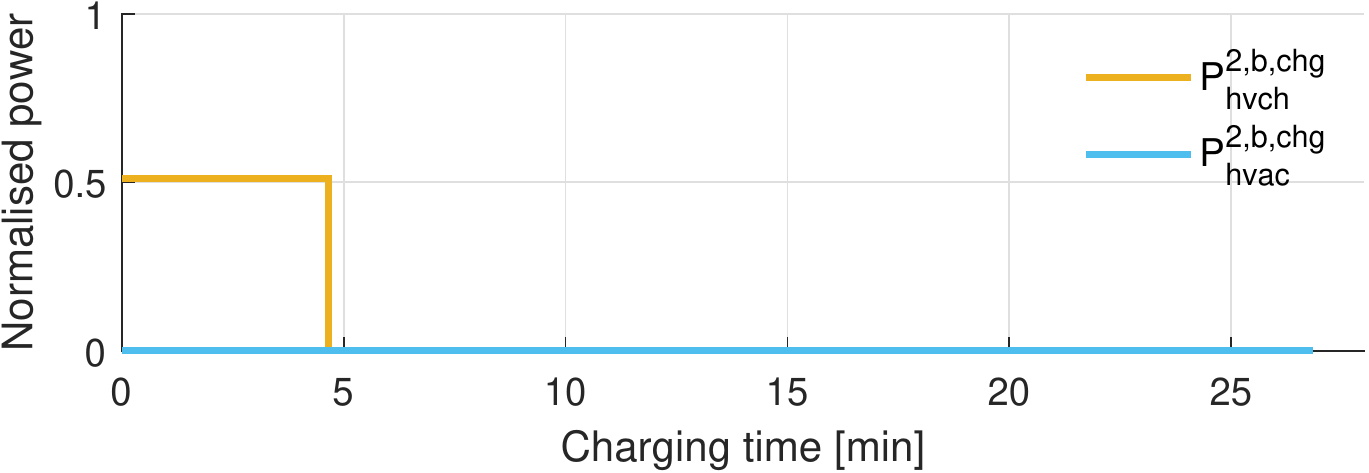}
\label{fig:caseC_hvchbhvacbchg2}
}
\caption{Case C: Energy optimal case with \SI{3}{kW} heat pump power limit.}
\label{fig:caseC}
\end{figure*}

Fig.~\ref{fig:caseC} demonstrates states and control inputs trajectories versus travelled distance and charging time. Changing the HP maximum power from \SI{1}{kW} to \SI{3}{kW} can considerably influence the energy optimal solution. Accordingly, only one charging stop is performed during the whole trip. Also, HVCH is used for battery heating before the charging stop and several minutes at the beginning of the charging period. As only the HP is used for cabin heating after the charging stop, the battery temperature is kept low and even drops below $\SI{0}{^\circ C}$ when approaching the destination. This combined with low SoC from the last \SI{50}{km} of the trip, results in a limited discharge power availability, which is a challenge for more aggressive driven cycles.

\subsection{Time Optimal Trip}\label{subsubsec:topt}
\begin{table}[t]
\caption{Time optimal solution at different ambient temperatures} \label{tab:topttab}
  \centering
  \setlength\tabcolsep{6pt}
 \begin{tabular}{|c|c @{\hspace{5.2ex}} c @{\hspace{5.2ex}} c|} 
 \hline
 \multicolumn{4}{| c |}{\textbf{-10 $^\circ$C ambient temperature}}\\
 \hline
 \begin{tabular}{c}Variable\end{tabular} & \begin{tabular}{c}HP disabled\end{tabular}& \begin{tabular}{c}smaller HP\end{tabular}&
 \begin{tabular}{c}larger HP\end{tabular}\\
 \hline
Time [min]& 25.9 & 24.2 & 23.5 \\
Reduction [\%] & - & 6.5 & 9.2 \\
\hline
\hline
 \multicolumn{4}{| c |}{\textbf{0 $^\circ$C ambient temperature}}\\
 \hline
 \begin{tabular}{c}Variable\end{tabular} & \begin{tabular}{c}HP disabled\end{tabular}& \begin{tabular}{c}smaller HP\end{tabular}&
 \begin{tabular}{c}larger HP\end{tabular}\\
 \hline
Time [min]& 22.3 & 15.5 & 15.5 \\
Reduction [\%] & - & 30.6 & 30.6 \\
\hline
\hline
 \multicolumn{4}{| c |}{\textbf{10 $^\circ$C ambient temperature}}\\
 \hline
 \begin{tabular}{c}Variable\end{tabular} & \begin{tabular}{c}HP disabled\end{tabular}& \begin{tabular}{c}smaller HP\end{tabular}&
 \begin{tabular}{c}larger HP\end{tabular}\\
 \hline
Time [min]& 15.8 & 14.2 & 14.2 \\
Reduction [\%] & - & 10.1 & 10.1 \\
\hline
\end{tabular}
\end{table}

Looking more closely at the time optimal solutions for different ambient temperatures illustrated in Fig.~\ref{fig:pf-10}-Fig.~\ref{fig:pf10}, reveals that the HP allows for shorter combined charging and detour time, compared to the case with HP disabled. For instance, the observed time reduction at \SI{-10}{\celsius} ambient temperature is \SI{6.5}{\%} and \SI{9.2}{\%} for the smaller and larger HP cases, respectively. Such time reduction is primarily due to a more efficient cabin heating during driving, leading to an improved grid-to-wheel efficiency; and thus reducing the amount of energy required to be supplied at a given charging stop. Furthermore, it may be infeasible to finish the trip with just one charging stop with the HP disabled. However, the number of charging stops can generally be reduced by having HP activated, which yields a lower total detour time. The detailed results about combined charging and detour time for different ambient temperatures and HP power limits are reported in Table \ref{tab:topttab}. In the following Section~\ref{subsec:case A}, the results of Case E, i.e. time optimal solution with \SI{1}{kW} HP power limit at \SI{-10}{^\circ C} ambient temperature, are demonstrated.

\subsubsection{Case E}\label{subsec:case E}
\begin{figure*}[t!]
\centering
\subfigure[Battery temperature and SoC trajectories vs. travelled distance.]{
 \includegraphics[width=.45\linewidth]{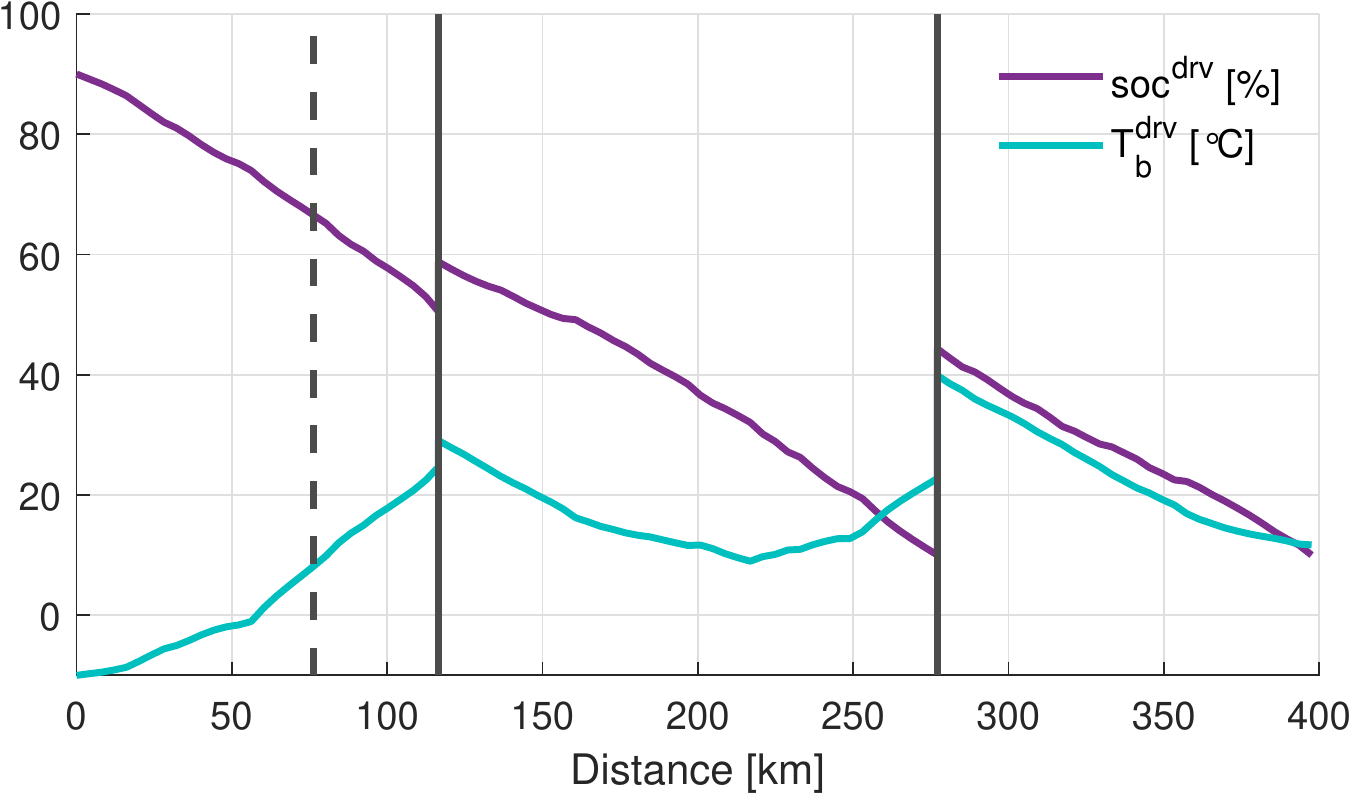}\hspace{.25cm}
\label{fig:caseE_soctbdrv}
}
\subfigure[Trajectories of battery power and propulsion power together with battery power limits vs. travelled distance.]{

 \includegraphics[width=.47\linewidth]{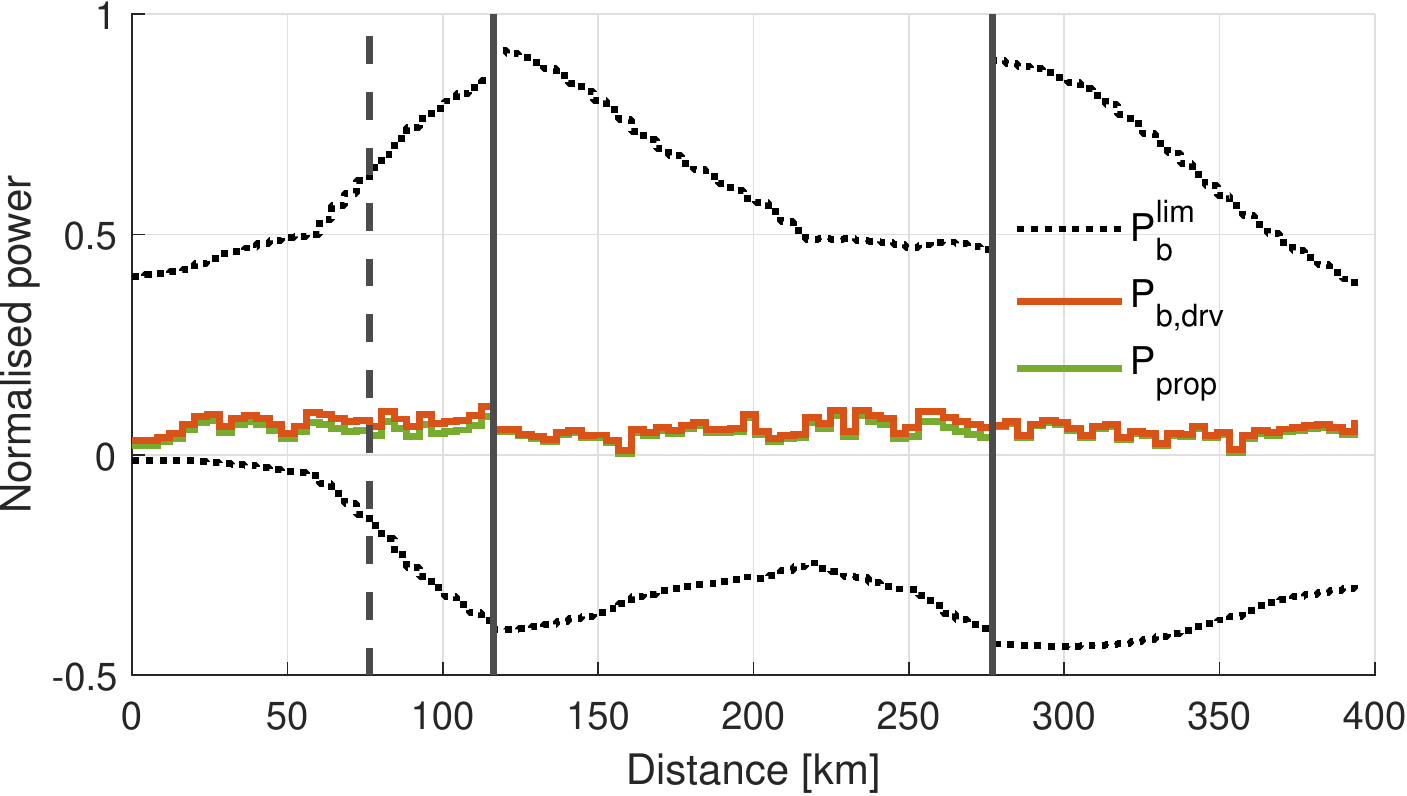}
\label{fig:caseE_pbpropdrv}
}
\subfigure[Trajectories of HVCH and HP power for cabin heating, vs. travelled distance.]{

 \includegraphics[width=.47\linewidth]{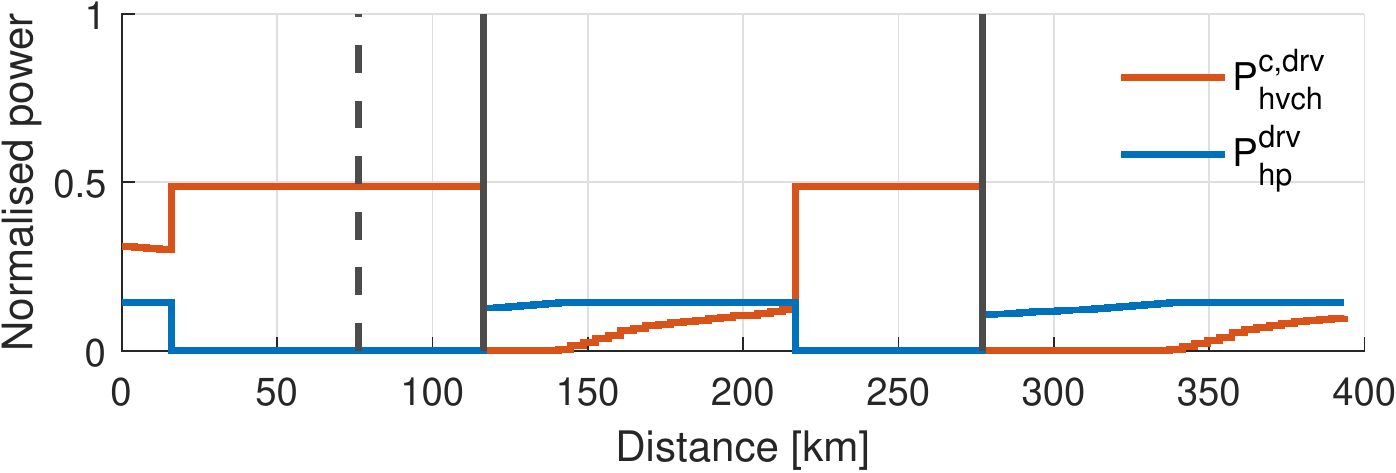}
\label{fig:caseE_hvchchpdrv}
}
\subfigure[Trajectories of HVCH and HVAC power, respectively for battery heating and cooling, vs. travelled distance.]{

 \includegraphics[width=.47\linewidth]{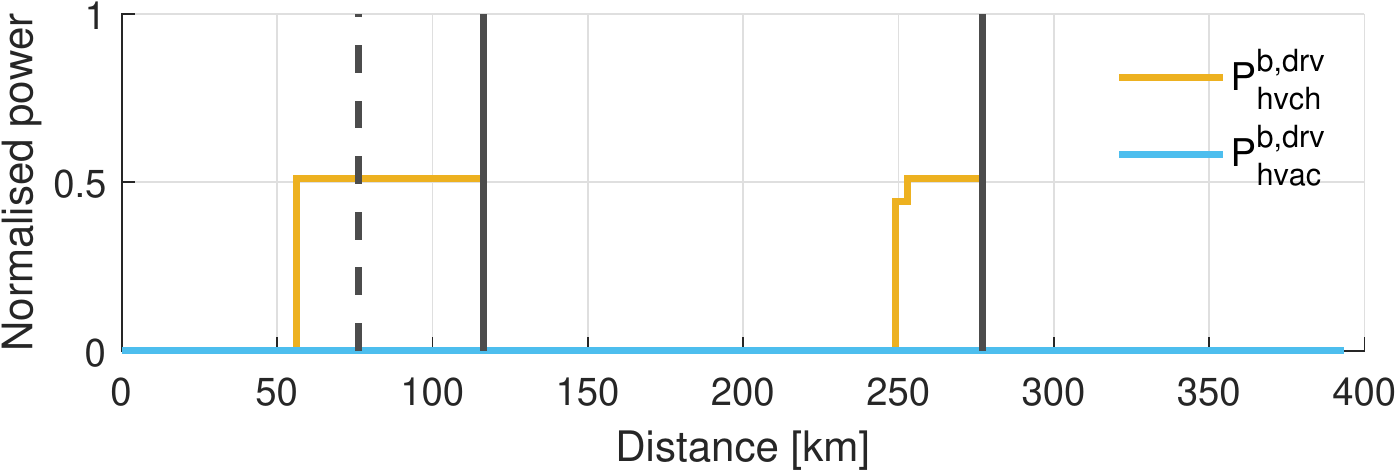}
\label{fig:caseE_hvchbhvacbdrv}
}
\subfigure[Battery temperature and SoC trajectories vs. charging time (second charging stop).]{

 \includegraphics[width=.47\linewidth]{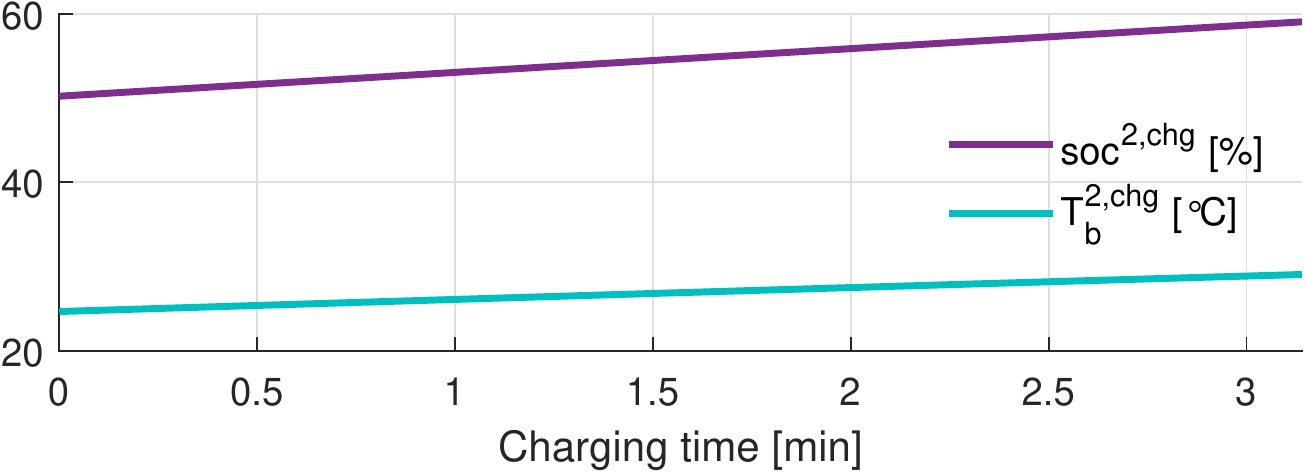}
\label{fig:caseE_soctbchg2}
}
\subfigure[Trajectories of HVCH and HP power for cabin heating, vs. charging time (second charging stop).]{

 \includegraphics[width=.47\linewidth]{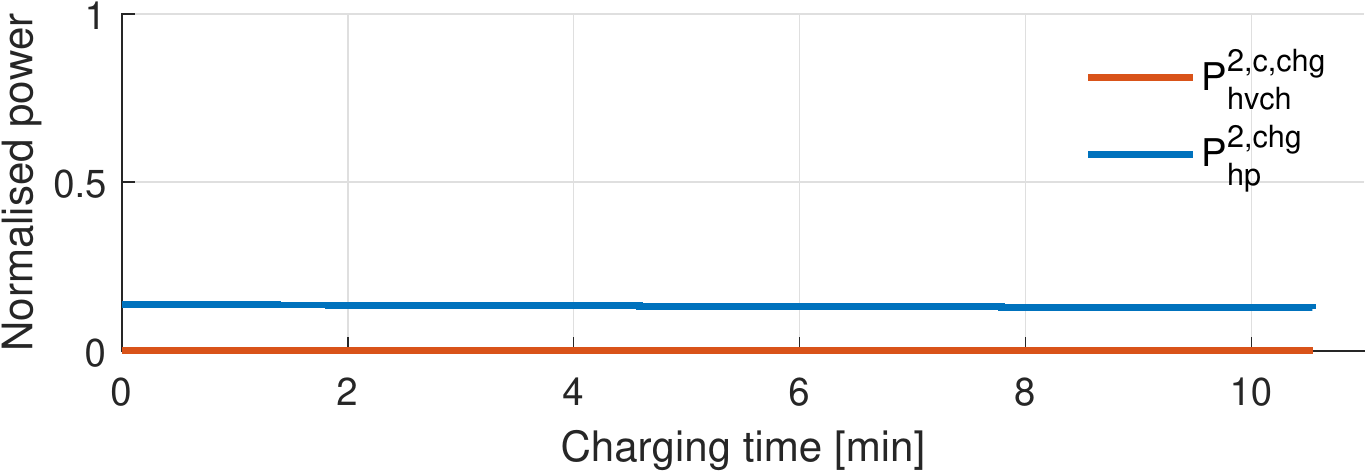}
\label{fig:caseE_hvchchpchg2}
}
\subfigure[Trajectories of HVCH and HVAC power, respectively for battery heating and cooling, vs. charging time (second charging stop).]{

 \includegraphics[width=.47\linewidth]{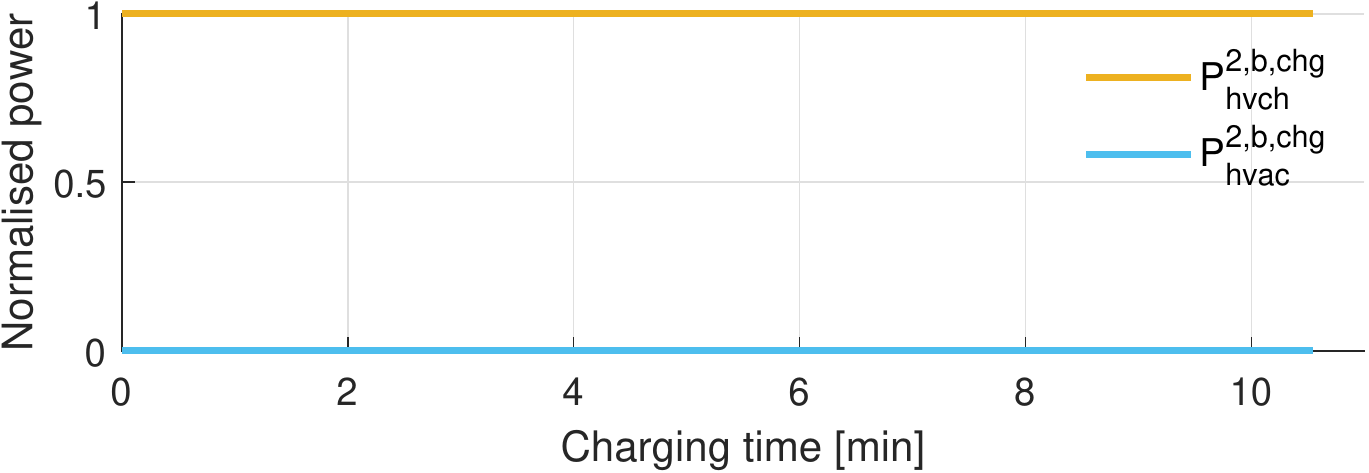}
\label{fig:caseE_hvchbhvacbchg2}
}
\subfigure[Battery temperature and SoC trajectories vs. charging time (third charging stop).]{

 \includegraphics[width=.47\linewidth]{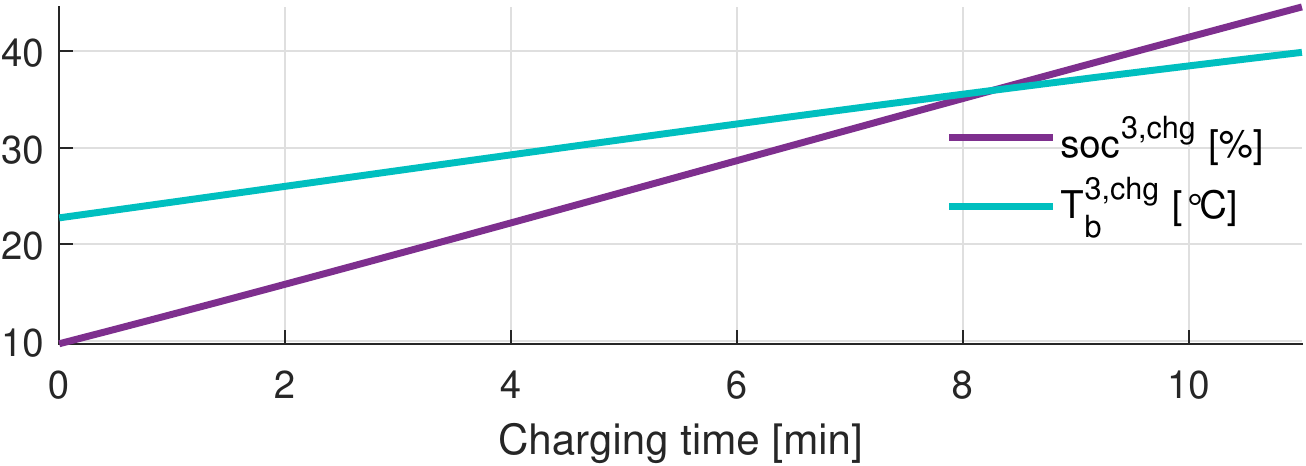}
\label{fig:caseE_soctbchg3}
}
\subfigure[Trajectories of HVCH and HP power for cabin heating, vs. charging time (third charging stop).]{

 \includegraphics[width=.47\linewidth]{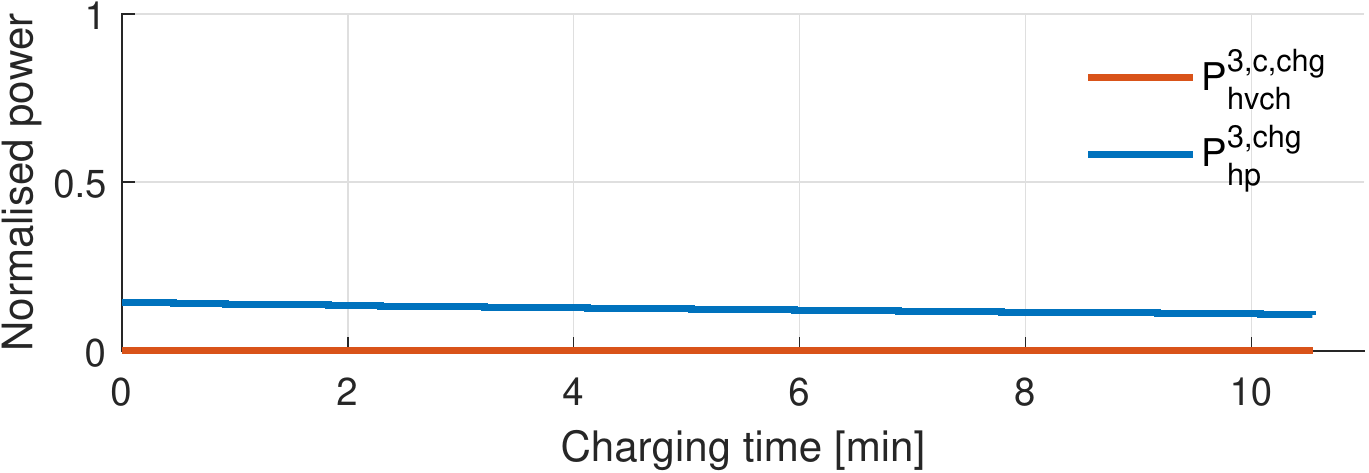}
\label{fig:caseE_hvchchpchg3}
}
\subfigure[Trajectories of HVCH and HVAC power, respectively for battery heating and cooling, vs. charging time (third charging stop).]{

 \includegraphics[width=.47\linewidth]{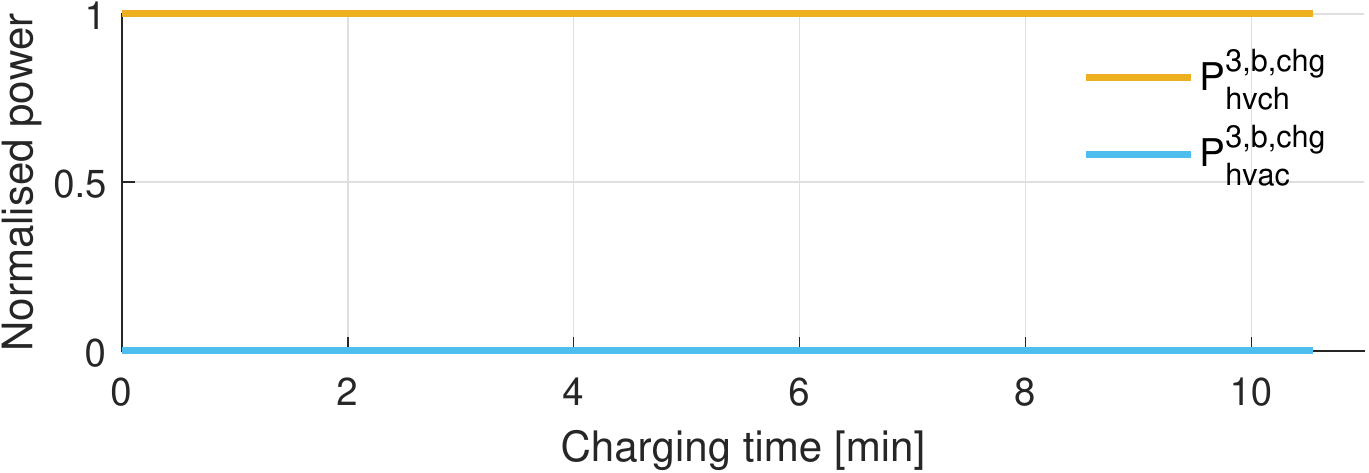}
\label{fig:caseE_hvchbhvacbchg3}
}
\caption{Case E: Time optimal case with \SI{1}{kW} heat pump power limit.}
\label{fig:caseE}
\end{figure*}

States and control inputs trajectories versus travelled distance and charging time are shown in Fig.~\ref{fig:caseE_soctbdrv}-Fig.~\ref{fig:caseE_hvchbhvacbchg3}. Similar to the energy optimal Case B, two charging stops are also performed in the time optimal Case E. However, in contrast to the energy optimal case, the battery is pre-heated before the charging stops, to the point with optimal temperature, i.e. $\approx \SI{25}{^\circ C}$, for fast charging. During charging, the use of HVCH at full power for battery heating and only HP for cabin heating at the same time is an effort to maximise the amount of heat possible to be within the battery pack. This implies that, the HP frees up the HVCH for maximum battery heating, while maintaining cabin heating demand.

% \subsection{Compromise between Energy and Time}\label{subsubsec:trade-off}
% To be written \dots

\subsection{Charged Energy vs. Ambient Temperature}\label{subsec:evstamb}
\begin{figure*}[t!]
\centering
\subfigure[Charged energy vs. ambient temperature for different maximum HP power values.]{
 \includegraphics[width=.425\linewidth]{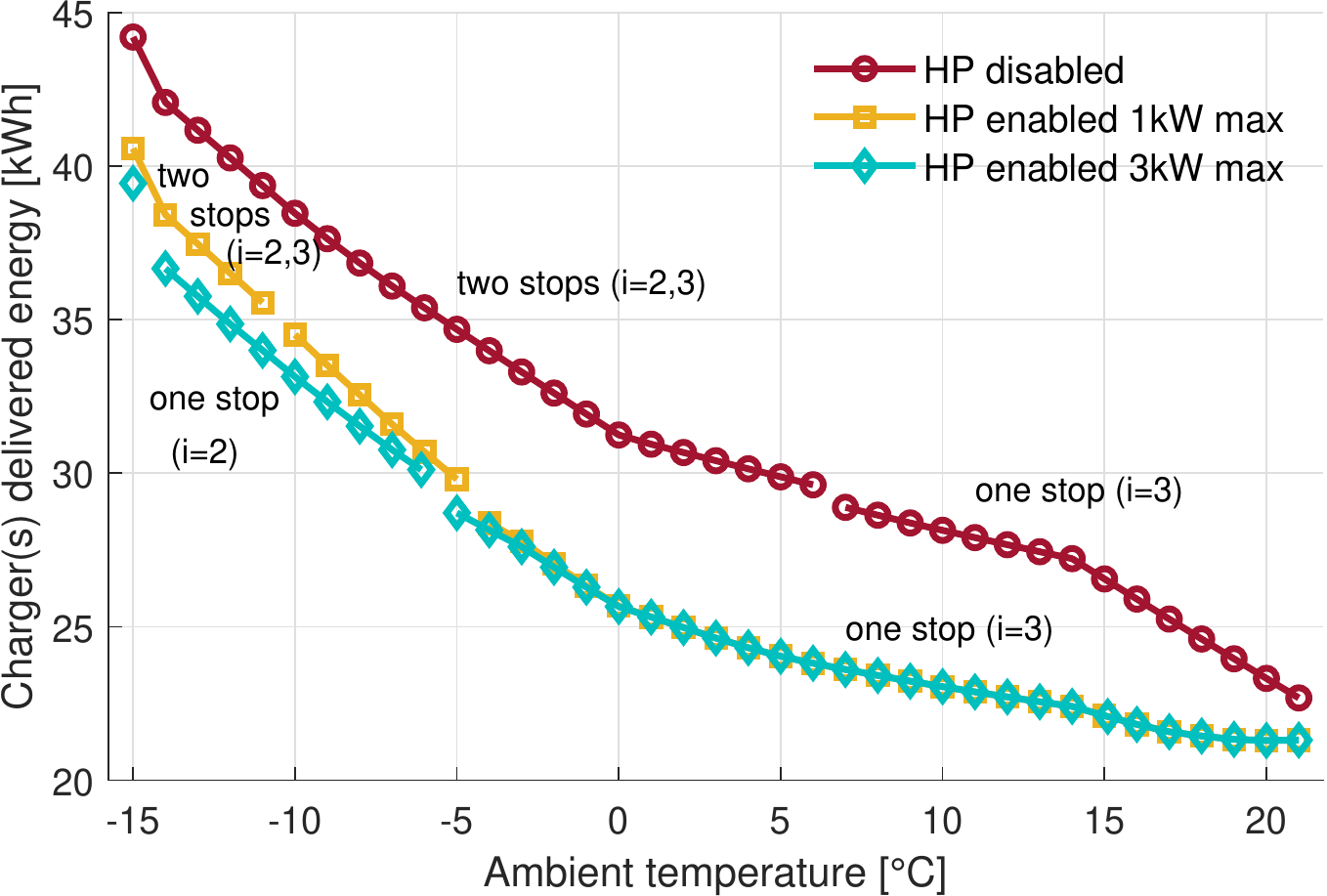}\hspace{1cm}
\label{fig:evstamba}
}
\subfigure[Relative energy benefit of HP activated cases compared to HP disabled case over ambient temperature.]{

 \includegraphics[width=.425\linewidth]{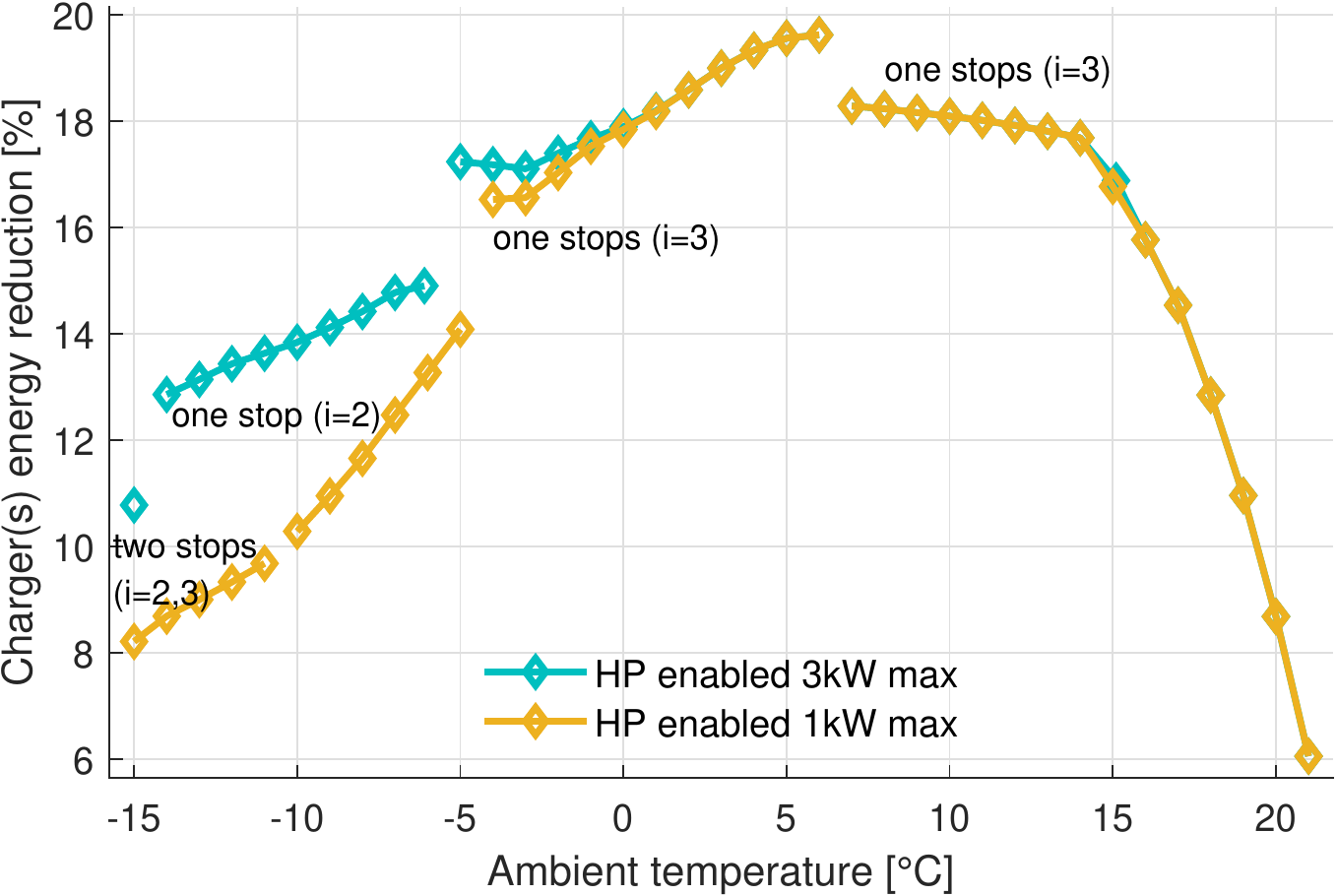}
\label{fig:evstambb}
}
\caption{Comparison of energy delivered by charger(s) over ambient temperature.}
\label{fig:evstamb}
\end{figure*}

Fig.~\ref{fig:evstamba} illustrates the charger(s) delivered energy versus ambient temperature values for different HP maximum power limits, and with the time cost fixed at \SI{40}{SEK/h}, which corresponds to point D in Fig.~\ref{fig:pf-10}. In Fig.~\ref{fig:evstamba}, number of charging stops for a given HP power limit and ambient temperature is also given. Accordingly, at high ambient temperatures between \SI{7}{^\circ C} and \SI{21}{^\circ C}, both the HP enabled and disabled cases are able to complete the trip with one late charging stop, i.e. ($i=3$). However, the HP enabled cases demand between \SI{6}{\%} to \SI{18}{\%} less energy from the charger compared to the HP disabled case, as depicted in Fig.~\ref{fig:evstambb}. At \SI{6}{^\circ C}, two charging stops are needed for the HP disabled case to complete the trip. Thus, a jump in energy reduction of about \SI{19.5}{\%} for the HP enabled cases is observed, which is due to the increased detour energy and time associated with stopping twice ($i=2,3$). As the ambient temperature is reduced further, the energy difference between the smaller and larger HP cases is more noticeable. This is due to a combination of high cabin heating demand and reduced CoP at low battery temperatures, resulting in the need for more than \SI{1}{kW} of HP compressor power to maintain the cabin climate. Thus, the more limited power case has to supplement cabin heating with the HVCH, while the other case is able to supplement less or not at all by the HVCH. Once the ambient temperature is dropped to \SI{-5}{^\circ C} and \SI{-6}{^\circ C}, the smaller and larger HP cases, respectively, start switching to perform one early stop at the second charge location ($i=2$). When this switch occurs, the energy reduction percentage drops, even though the detour time or detour energy has not changed. This is due to the low charging power in the high SoC region, which leads to a longer charging time; and accordingly an increased energy demand by the auxiliary and TM system components. With the temperature reduced to \SI{-11}{^\circ C}, the smaller HP case starts to perform two charging stops, the same way as the HP disabled case. Such switch occurs for the \SI{3}{kW} limit case at \SI{-15}{^\circ C}.

\section{Discussion}\label{sec:dis}
Here the benefits of including HP in the TM system and optimally planning the charging points are discussed.

\subsection{Improved Energy Efficiency and Trip Time by a Heat Pump}\label{subsec:hpinc}
According to the results given in Section~\ref{sec:res}, the reduction in terms of both energy consumption as well as combined charging and detour time is significant, when an HP is considered in the TM system of BEVs for waste heat recovery. Although the improvement varies noticeably with ambient temperature, as long as there is a heating demand for the cabin compartment, the case with an HP activated has better performance compared to the one without. 
This is true even when the HP compressor power is limited, especially in milder climates.

Using an HP in the TM system may be less advantageous in cases where the waste heat available within the battery pack is limited, or there are constraints on discharge power capability of the battery at low SoC and temperature regions.

\subsection{Effects of Charge Point Optimisation}\label{subsec:chgopt}
Optimal charge point planning allows for a holistic solution of a long trip in a BEV in terms of energy consumption and total trip time. At warmer ambient temperatures, i.e. $T_\tx{amb}\geq\SI{0}{^\circ C}$, minimum possible number of charging stops is favourable, regardless of priorities in terms of time or energy, suggested by Fig.~\ref{fig:pf0} and Fig.~\ref{fig:pf10}. On the other hand, at colder ambient temperatures, e.g. $T_\tx{amb}=\SI{-10}{^\circ C}$, two charging stops are identified to be energy and/or time optimal for HP disabled and smaller HP, as shown in Fig.~\ref{fig:pf-10}. This implies that the increased consumption due to higher demand for cabin heating outweighs the energy and time cost associated with stopping frequently. Thus, there is a merit to the strategy of initially driving as far as possible to a stop, in which charging is performed enough to make it to the next charging station. However, as demonstrated by the trade-off between the two extremes, energy and time optimal, with the \SI{1}{kW} limit, there are cases where that strategy is not the optimal solution. For e.g. in Case D only one charging stop is performed, where the optimal strategy suggests minimising the detour energy and time by reducing the number of stops.

\section{Conclusion and Future Work}\label{sec:con}
In this paper, a mixed-integer nonlinear optimisation problem is formulated for optimal thermal management and charging of a BEV, in order to capture its long trip including both driving and charging. Within this problem, Pareto frontiers describing the trade-off between energy efficiency and time are derived versus different features, e.g. a heat pump, charging stops, and ambient temperature. Such graphs provide a wide range of choices for car manufacturers as well as grid service providers to gain more insight into the design and development of TM and charging systems. Furthermore, various car users can customise their trips according to the information given within these graphs. According to the obtained results, energy consumption and the time needed for charging are reduced by up to \SI{19.4}{\%} and \SI{30.6}{\%}, respectively, by including an HP in the TM system. By including optimal charge point planning in the form of binary decision variables, the solution depends on factors such as the priority between time and energy, the availability of an HP, and ambient temperature. % characterised as the cabin heating demand.

The current study can readily be extended by the inclusion of speed optimisation in favour of energy-efficient driving, so-called \textit{eco-driving}, where the vehicle's longitudinal dynamics is required to be incorporated in the problem formulation, in addition to the dynamics of battery temperature and SoC. Note that a similar analysis has been conducted in~\cite{hamednia2022a}, but without charge point planning and without an HP. Thus, a nonuniform sampling could be introduced, or speed could be optimized on a separate level. With such an extension in the developed algorithm, the solution would represent a more complete route optimisation, aiming at enhancing energy and/or time efficiency. For instance, the short charging periods at the second location in Case B and Case E may be avoided if the vehicle eco-drives, leading to a direct reduction in time and energy.

In order to implement the proposed algorithm online in a vehicle, it is crucial to reduce the computational burden. To do so, the knowledge gained by the current results is highly beneficial. For instance, in the case of active battery pre-heating, the solution always involves running the HVCH at maximum power for some period right before the charging stop and at the beginning of the charging interval. According to such knowledge, one effort may be to re-formulate the problem to control the average power or energy used for battery heating instead. This may allow for a significant reduction in discretised samples to imitate the non-simplified system behaviour, with a small or non-existent loss in optimality.

\appendices

\section*{Acknowledgment}
The authors would like to acknowledge Mats Bohman from Volvo Car Corporation for the support and fruitful discussions during this research. This work is part of a project titled ``Predictive Energy and Thermal management of Electric Vehicles with Connectivity to Infrastructure" funded by the Swedish Electromobility Center.

\nocite{lopez17}
\bibliographystyle{IEEEtran}
\bibliography{bibliography_ah}

\end{document}